\documentclass[sigconf]{acmart}

\usepackage{multirow} 
\newcommand{\tabincell}[2]{\begin{tabular}{@{}#1@{}}#2\end{tabular}}

\usepackage{subfigure} 

\usepackage{bm}
\usepackage{url}
\usepackage{hyperref}
\hypersetup{
    colorlinks=true,
    linkcolor=blue,
    filecolor=magenta,      
    urlcolor=cyan,
}

\usepackage{listings}
\usepackage{xcolor}
\definecolor{codegreen}{rgb}{0,0.6,0}
\definecolor{codegray}{rgb}{0.5,0.5,0.5}
\definecolor{codepurple}{rgb}{0.58,0,0.82}
\definecolor{backcolour}{rgb}{0.95,0.95,0.92}
\lstdefinestyle{mystyle}{
    backgroundcolor=\color{backcolour},   
    commentstyle=\color{codegreen},
    keywordstyle=\color{blue},  
    numberstyle=\tiny\color{codegray},
    stringstyle=\color{codepurple},
    basicstyle=\ttfamily\footnotesize,
    breakatwhitespace=false,
    breaklines=true,
    captionpos=b, 
    keepspaces=true,
    numbers=left, 
    numbersep=5pt, 
    showspaces=false,
    showstringspaces=false,
    showtabs=false, 
    tabsize=2
}
\usepackage{algorithm}
\usepackage{algorithmic}
\usepackage{setspace}

\usepackage{pifont}

\newcommand{\toolname}{TranCS}

\AtBeginDocument{%
  \providecommand\BibTeX{{%
    \normalfont B\kern-0.5em{\scshape i\kern-0.25em b}\kern-0.8em\TeX}}}

\copyrightyear{2022}
\acmYear{2022}
\setcopyright{acmcopyright}
\acmConference[ICSE '22]{44th International Conference on Software Engineering}{May 21--29, 2022}{Pittsburgh, PA, USA} 
\acmBooktitle{44th International Conference on Software Engineering (ICSE '22), May 21--29, 2022, Pittsburgh, P A, USA}
\acmPrice{15.00}
\acmDOI{10.1145/3510003.3510140}
\acmISBN{978-1-4503-9221-1/22/05}

\begin{document}

\title{Code Search based on Context-aware Code Translation}

\author{Weisong Sun}
\email{weisongsun@smail.nju.edu.cn}
\affiliation{%
  \institution{State Key Laboratory for Novel Software Technology}
  \city{Nanjing University}
  \country{China}
}

\author{Chunrong Fang}
\authornote{Corresponding author.}
\email{fangchunrong@nju.edu.cn}
\affiliation{%
  \institution{State Key Laboratory for Novel Software Technology}
  \city{Nanjing University}
  \country{China}
}

\author{Yuchen Chen}
\email{yuc.chen@outlook.com}
\affiliation{%
  \institution{State Key Laboratory for Novel Software Technology}
  \city{Nanjing University}
  \country{China}
}

\author{Guanhong Tao}
\email{taog@purdue.edu}
\affiliation{%
  \institution{Purdue University}
  \city{West Lafayette}
  \state{Indiana}
  \country{USA}
}

\author{Tingxu Han}
\email{hantingxv@163.com}
\affiliation{%
  \institution{School of Information Management}
  \city{Nanjing University}
  \country{China}
}

\author{Quanjun Zhang}
\email{quanjun.zhang@smail.nju.edu.cn}
\affiliation{%
  \institution{State Key Laboratory for Novel Software Technology}
  \city{Nanjing University}
  \country{China}
}

\renewcommand{\shortauthors}{Weisong Sun and Chunrong Fang, et al.}

\begin{abstract}
Code search is a widely used technique by developers during software development. 
It provides semantically similar implementations from a large code corpus to developers based on their queries. 
Existing techniques leverage deep learning models to construct embedding representations for code snippets and queries, respectively. 
Features such as abstract syntactic trees, control flow graphs, etc., are commonly employed for representing the semantics of code snippets. 
However, the same structure of these features does not necessarily denote the same semantics of code snippets, and vice versa. 
In addition, these techniques utilize multiple different word mapping functions that map query words/code tokens to embedding representations. 
This causes diverged embeddings of the same word/token in queries and code snippets. 
We propose a novel context-aware code translation technique that translates code snippets into natural language descriptions (called translations). 
The code translation is conducted on machine instructions, where the context information is collected by simulating the execution of instructions. 
We further design a shared word mapping function using one single vocabulary for generating embeddings for both translations and queries. 
We evaluate the effectiveness of our technique, called {\toolname}, on the CodeSearchNet corpus with 1,000 queries. 
Experimental results show that {\toolname} significantly outperforms state-of-the-art techniques by 49.31\% to 66.50\% in terms of $\mathtt{MRR}$ (mean reciprocal rank).
\end{abstract}



\maketitle

\section{Introduction}
\label{sec:introduction}
\begin{figure}[htbp]
  \centering
  \includegraphics[width=0.9\linewidth]{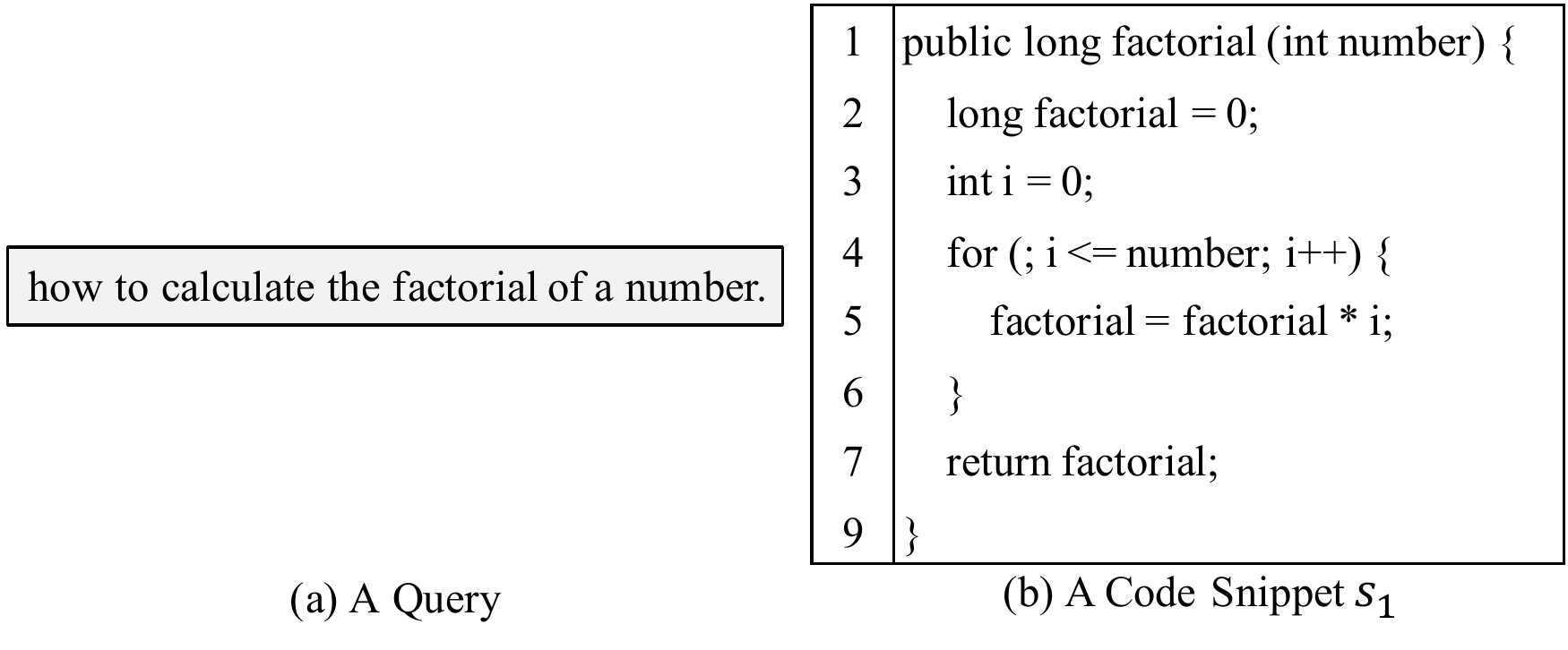}
  \caption{An Example of Query and Code Snippet}
  \label{fig:example_of_query_and_code_snippet}
\end{figure}

Software development is usually a repetitive task, where same or similar implementations exist in established projects or online forums. 
Developers tend to search for those high-quality implementations for reference or reuse, so as to enhance the productivity and quality of their development~\cite{2009-Two-Studies-of-Opportunistic-Programming, 2010-Example-centric-programming, 2017-Cross-project-Code-Reuse}. 
Existing studies~\cite{2009-Two-Studies-of-Opportunistic-Programming, 2020-Code-Search-with-Co-Attentive-Representation} show that developers often spend 19\% of their time on finding reusable code examples during software development.
Code search (CS) is an active research field~\cite{1997-Examination-of-SE-Work-Practices, 2012-Recommending-Source-Code, 2015-How-Developers-Search-for-Code, 2017-What-do-Developer-Search-for, 2018-DeepCodeSearch, 2019-DL-Met-CodeSearch, 2019-Multi-modal-Attention-for-Code-Rerieval, 2020-Code-Search-with-Co-Attentive-Representation, 2021-Two-Stage-Attention-for-Code-Search, 2021-deGraphCS}, which aims at designing advanced techniques to support code retrieval services.
Given a query by the developer, CS retrieves code snippets that are related to the query from a large-scale code corpus, such as GitHub~\cite{2008-GtHub} and Stack Overflow~\cite{2008-Stack-Overflow}. 
Figure~\ref{fig:example_of_query_and_code_snippet} shows an example. 
The query ``how to calculate the factorial of a number'' in Figure~\ref{fig:example_of_query_and_code_snippet}(a) is provided by the developer, which is usually a short natural language sentence describing the functionality of the desired code snippet~\cite{2020-Opportunities-in-Code-Search}. 
The method/function~\cite{2020-Code-Search-with-Co-Attentive-Representation, 2019-Multi-modal-Attention-for-Code-Rerieval, 2016-Query-Expansion-based-Crowd-Knowledge, 2014-Spotting-Working-Code} in Figure~\ref{fig:example_of_query_and_code_snippet}(b) is a possible code snippet that satisfies the developer's requirement.

Existing CS techniques can be categorized into \textit{traditional methods} that use keyword matching between queries and code snippets such as information retrieval-based code search~\cite{2006-Source-Code-Exploration, 2010-Example-centric-programming, 2011-Portfolio, 2014-Spotting-Working-Code, 2015-How-Developers-Search-for-Code} and query reformulation-based code search~\cite{2009-Automatically-Capture-Code-Context, 2014-Thesaurus-based-Query-Expansion, 2015-CodeHow, 2015-Query-Expansion-via-WordNet, 2016-Query-Expansion-based-Crowd-Knowledge, 2018-Effective-Reformulation-of-Query}, and \textit{deep learning methods} that encode queries and code snippets into embedding representations capturing semantic information. 
Traditional methods simply treat queries and code snippets as plain texts, and retrieve query-related code snippets by only looking at matched keywords. 
They fail to capture the semantics of both query sentences and code snippets. 
Deep learning (DL) methods transform input queries and code snippets into embedding representations. 
Specifically, for a given query, all the words in the query sentence are first represented as word embeddings and then fed to a DL model to produce a query embedding~\cite{2018-DeepCodeSearch, 2020-Code-Search-with-Co-Attentive-Representation}. 
For a code snippet, multiple aspects are extracted as features, such as tokens, abstract syntactic trees (ASTs), and control flow graphs (CFGs). 
These features are transformed into corresponding embeddings and processed by another DL model to produce a code embedding~\cite{2018-DeepCodeSearch, 2020-Code-Search-with-Co-Attentive-Representation, 2020-Functional-Code-Clone-Detection, 2021-Two-Stage-Attention-for-Code-Search, 2021-deGraphCS}. 
The code search task is hence to find similar pairs between query embeddings and code embeddings. 
While \textit{DL methods} surpass \textit{traditional methods} in capturing the semantics of queries and code snippets, their performances are still limited due to the insufficiency of encoding semantics and the embedding discrepancy between queries and code snippets. 
Existing techniques miss either data dependencies among code statements like MMAN~ \cite{2019-Multi-modal-Attention-for-Code-Rerieval} or control dependencies such as DeepCS~\cite{2018-DeepCodeSearch}, CARLCS-CNN \cite{2020-Code-Search-with-Co-Attentive-Representation}, and TabCS \cite{2021-Two-Stage-Attention-for-Code-Search}. 
Furthermore, the embedding representations of code snippets are largely different from those of query sentences written in natural language, causing semantic mismatch during the code search task. 
For example, MMAN~\cite{2019-Multi-modal-Attention-for-Code-Rerieval} uses different word mapping functions (that map a word or token to an embedding representation) to encode queries, and tokens, ASTs, and CFGs in code snippets. 
For the widely used word \texttt{length} in both queries and code snippets, the embedding representations are different in those word mapping functions, leading to poor code search performance as we will discuss in Section~\ref{sec:motivation} and experimentally show in Section~\ref{subsubsec:contribution_of_each_component}.

We propose a novel context-aware code translation technique that translates code snippets into natural language descriptions (called translations). 
Such a translation can bridge the representation discrepancy between code snippets (in programming languages) and queries (in natural language). 
Specifically, we utilize a standard program compiler and a disassembler to generate the instruction sequence of a code snippet. However, the context information such as local variables, data dependency, etc., are missed from the instruction sequence. 
We hence simulates the execution of instructions to collect those desired contexts. 
A set of pre-defined translation rules are then used to translate the instruction sequence and contexts into translations. 
Such a code translation is context-aware. 
The translations of code snippets are similar to those descriptions in queries, in which they share a range of words. 
We hence design a shared word mapping mechanism using one single vocabulary for generating embeddings for both translations and queries, substantially reducing the semantic discrepancy and improving the overall performance (see results in Section~\ref{subsubsec:contribution_of_each_component}).

In summary, we make the following contributions.
\begin{itemize}
	\item We propose a context-aware code translation technique that transforms code snippets into natural language descriptions with preserved semantics.

	\item We introduce a shared word mapping mechanism, which bridges the discrepancy of embedding representations from code snippets and queries.

	\item We implement a code search prototype called {\toolname}. We evaluate it on the CodeSearchNet corpus~\cite{2019-CodeSearchNet-Challenge} with 1,000 queries. Experimental results show that {\toolname} improves the top-1 hit rate of code search by 67.16\% to 102.90\% compared to state-of-the-art techniques. In addition, {\toolname} achieves $\mathtt{MRR}$ of 0.651, outperforming DeepCS~\cite{2018-DeepCodeSearch} and MMAN~\cite{2019-Multi-modal-Attention-for-Code-Rerieval} by 66.50\% and 49.31\%, respectively. The source code of {\toolname} and all the data used in this paper are released and can be downloaded from the website~\cite{2021-TranCS}.
\end{itemize}

\section{Background}
\label{sec:background}

\subsection{Machine Instruction}
\label{subsec:machine_instruction}
Since the context-aware code translation technique we propose is performed at the machine instruction level, we first introduce the background about machine instructions. 

A program runs by executing a sequence of machine instructions~\cite{2014-CFG-based-Opcode-Analysis-Malware}. A machine instruction consists of an opcode specifying the operation to be performed, followed by zero or more operands embodying values to be operated upon~\cite{2008-Unknown-Malcode-Detection-Using-Opcode, 2021-JVM-Specification}. For example, in Java Virtual Machine, $\mathtt{istore\_2}$ is a machine instruction where $\mathtt{istore}$ is an opcode whose operation is ``store $\mathtt{int}$ into local variable", and $\mathtt{2}$ is an operand that represents the index of the local variable. Machine instructions have been widely used in software engineering activities, such as malware detection~\cite{2007-Opcodes-as-predictor-for-malware, 2008-Unknown-Malcode-Detection-Using-Opcode, 2014-CFG-based-Opcode-Analysis-Malware}, API recommendation~\cite{2016-Learning-API-Usages-from-Bytecode}, code clone detection~\cite{2018-DL-Similarities-from-Code-Representations}, program repair~\cite{2019-Program-Repair-via-Bytecode-Mutation}, and binary code search~\cite{2019-Cross-OS-Binary-Code-Search}. Machine instructions are generated by disassembling the binary files, such as the $\mathtt{.class}$ file in Java. Therefore, it is also called bytecode~\cite{2016-Learning-API-Usages-from-Bytecode, 2016-Code-Relatives, 2019-Program-Repair-via-Bytecode-Mutation} or bytecode mnemonic opcode~\cite{2018-DL-Similarities-from-Code-Representations} in some of the works mentioned above. For ease of understanding, the terminology ``instruction'' is used uniformly in this paper. 

\subsection{Deep Learning-based Code Search}
\label{subsec:dl-based_code_search}

\begin{figure}[htbp]
  \centering
  \includegraphics[width=\linewidth]{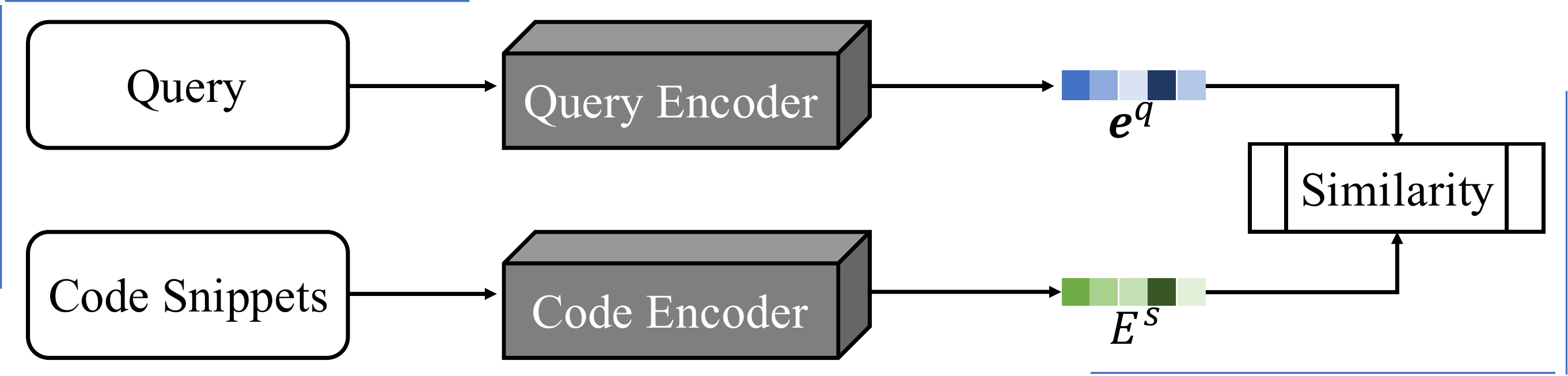}
  \caption{A General Framework of DL-based CS techniques}
  \label{fig:DL-based_CS}
\end{figure}

As shown in Figure \ref{fig:DL-based_CS}, we can observe that deep learning (DL)-based CS techniques usually consist of three components, a query encoder, a code encoder, and a similarity measurement component. The query encoder is an embedding network that can encode the query $q$ given by the developer into a $d$-dimensional embedding representation $\bm{e}^q \in \mathbb{R}^d$. To train such a query encoder, existing DL-based CS techniques have tried various neural network architectures, such as RNN~\cite{2018-DeepCodeSearch}, LSTM~\cite{2019-Multi-modal-Attention-for-Code-Rerieval}, and CNN~\cite{2021-Two-Stage-Attention-for-Code-Search}. In DL-based CS studies, it is a common practice to use code comments as queries during the training phase of the encoder~\cite{2018-DeepCodeSearch, 2019-Multi-modal-Attention-for-Code-Rerieval, 2020-Code-Search-with-Co-Attentive-Representation}. Code comments are natural language descriptions used to explain what the code snippets want to do~\cite{2018-Deep-Code-Comment-Generation}. For example, the first line of Figure~\ref{fig:example_of_code_snippets_with_same_semantic}(a) is a comment for the code snippet $s_a$. Therefore, we do not strictly distinguish the meaning of the two terms \textit{comment} and \textit{query}, and use the term \textit{comment} during encoder training, and \textit{query} at other times. The code encoder is also an embedding network that can encode $n$ code snippets in the code corpus $S$ into corresponding embedding representations $\bm{E}^S \in \mathbb{R}^{n\times d}$. In existing DL-based CS techniques, the code encoder is usually much more complicated than the query encoder. For example, the code encoder of MMAN~\cite{2019-Multi-modal-Attention-for-Code-Rerieval} consists of three sub-encoders that are built on the LSTM~\cite{1997-LSTM}, Tree-LSTM~\cite{2015-Tree-LSTM}, and GGNN~\cite{2016-GGNN} architectures with the goal of encoding different features of the code snippet, e.g., tokens, ASTs, and CFGs. The similarity measurement component is used to measure the cosine similarity between $\bm{e}^q$ and each $\bm{e}^s \in \bm{E}^s$. The target of DL-based CS techniques is to rank all code snippets in $S$ by the cosine similarity~\cite{2018-DeepCodeSearch}. The higher the similarity, the higher relevance of the code snippet to the given query.

\section{Motivation}
\label{sec:motivation}
In this section, we study the limitations of commonly used representations of code snippets as well as the representation discrepancy between code snippets and comments in existing works~\cite{2018-DeepCodeSearch, 2019-Multi-modal-Attention-for-Code-Rerieval}. 

\begin{figure}[htbp]
  \centering
  \includegraphics[width=\linewidth]{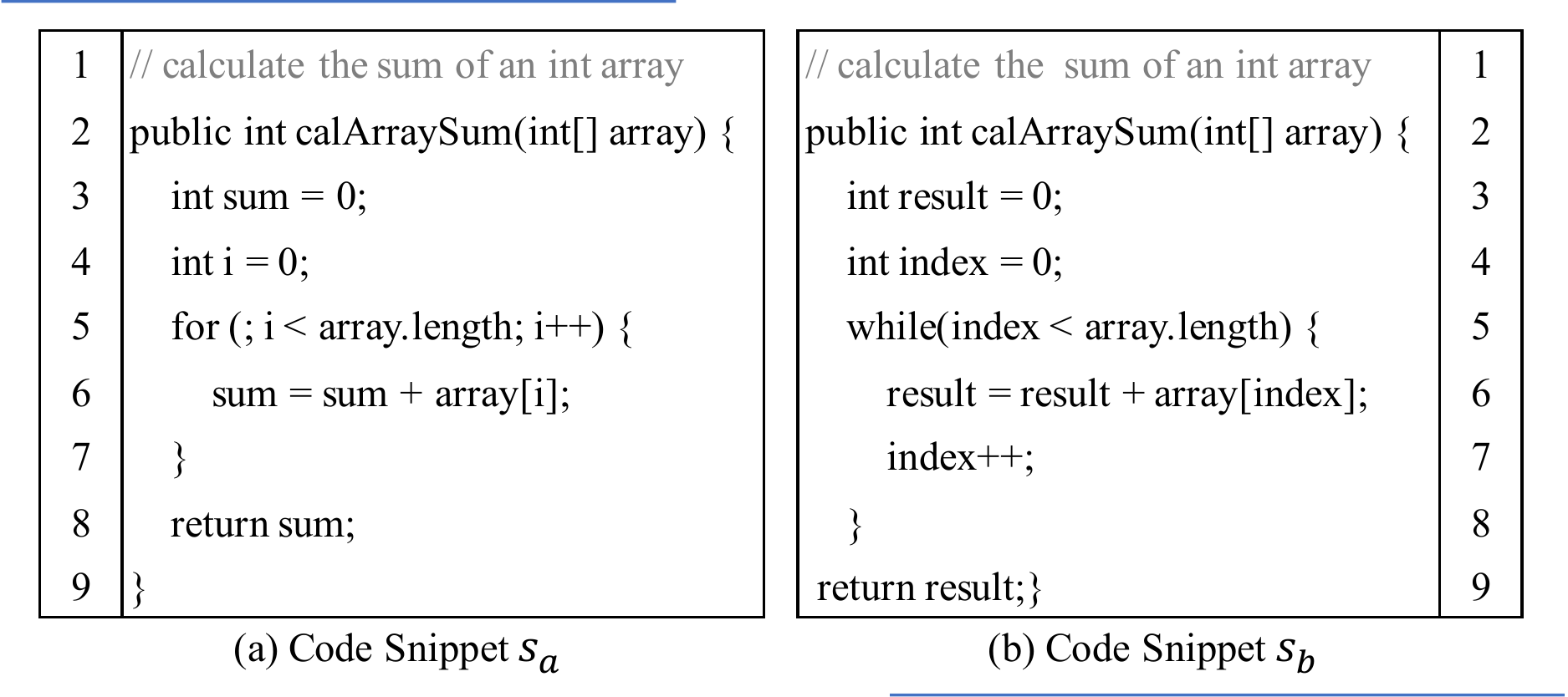}
  \caption{Code Snippets}
  \label{fig:example_of_code_snippets_with_same_semantic}
\end{figure}

\begin{figure}[htbp]
  \centering
  \includegraphics[width=\linewidth]{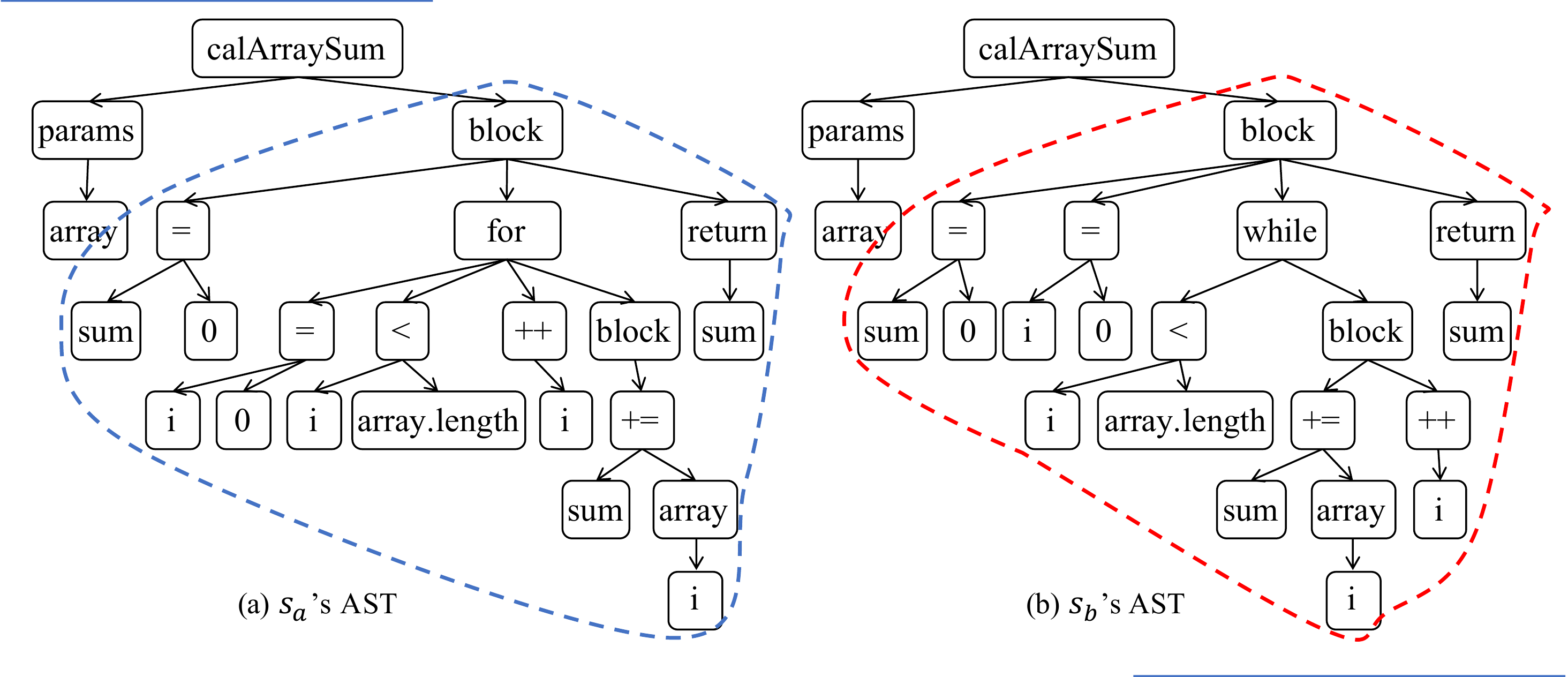}
  \caption{Abstract Syntactic Trees}
  \label{fig:example_of_ast}
\end{figure} 

\begin{figure}[!t]
  \centering
  \includegraphics[width=\linewidth]{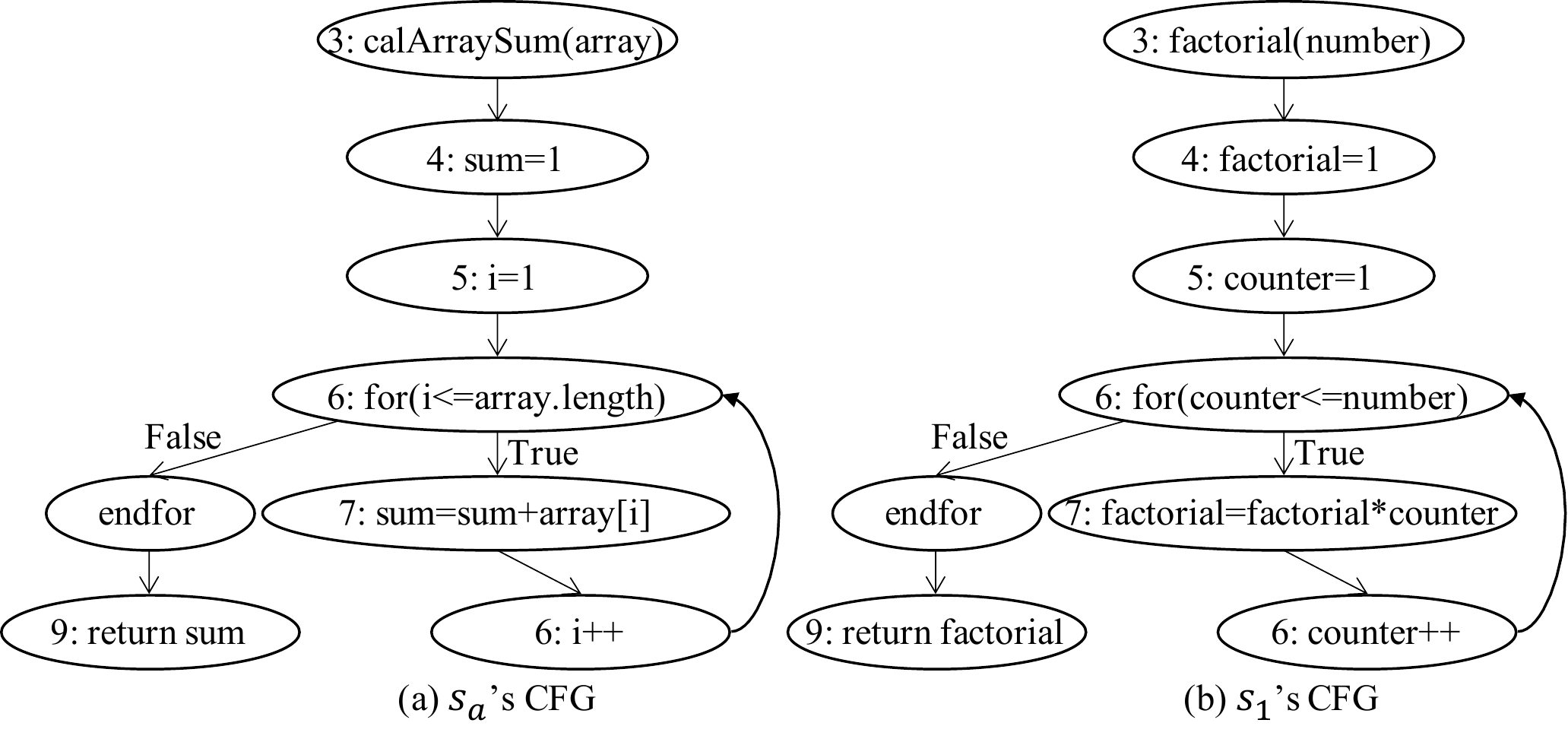}
  \caption{Control Flow Graphs}
  \label{fig:example_of_cfg}
\end{figure}

Figure~\ref{fig:example_of_code_snippets_with_same_semantic} shows two code snippets for calculating the sum of a given \texttt{int} array. Figure~\ref{fig:example_of_code_snippets_with_same_semantic}(a) uses a \texttt{for} statement to loop over all the elements in the array (line 5) and add their values to variable \texttt{sum} (line 6). Figure~\ref{fig:example_of_code_snippets_with_same_semantic}(b) employs a \texttt{while} statement for the same task (lines 5-8). Semantically, the two code snippets have the exact same meaning. In Figure~\ref{fig:example_of_ast}, we show the abstract syntax trees (ASTs) for the above two code snippets $s_a$ (left figure) and $s_b$ (right figure), respectively. Observe that the sub-trees circled in dotted lines are different for the two code snippets. Such representations cause the inconsistency of code semantics, leading to inferior results in code search as we will show in Section~\ref{subsubsec:effectiveness_of_TranCS}. Control flow graph (CFG) is also commonly used for representing code snippets. Figure~\ref{fig:example_of_cfg} depicts the CFGs for the two code snippets $s_a$ and $s_1$ (see Figure~\ref{fig:example_of_query_and_code_snippet}(b) in Section~\ref{sec:introduction}). The task of $s_1$ is to calculate the factorial of a given number, while $s_a$ is to calculate the sum of a given array. The two code snippets have completely different goals. However, the CFGs shown in Figure~\ref{fig:example_of_cfg} have the same graph structure, which cannot differentiate the semantic difference between the two code snippets. This example delineates the insufficiency of utilizing CFGs for representing code semantics. Our experimental results in Section~\ref{subsubsec:effectiveness_of_TranCS} show that a state-of-the-art technique MMAN~\cite{2019-Multi-modal-Attention-for-Code-Rerieval} leveraging ASTs and CFGs has a limited performance. 

Existing techniques leverage deep learning models (i.e., the encoders introduced in Section~\ref{subsec:dl-based_code_search}) for code search, where code snippets and comments need to be transformed into numerical forms in order to train those models and produce desired outputs. A common way is to build vocabularies for code snippets and comments, and construct corresponding numerical representations (e.g., word embeddings). A word mapping function is a dictionary with the key of a token in code snippets or a word in comments (from vocabularies) and the value of a fixed-length real-valued vector. DeepCS~\cite{2018-DeepCodeSearch} builds four mapping functions for method names (MN), API sequences (APIs), tokens, and comments, separately. MMAN~\cite{2019-Multi-modal-Attention-for-Code-Rerieval} utilizes four different mapping functions for tokens, ASTs, CFGs, and comments, respectively. The embeddings in these mapping functions are randomly initialized and learned during the training process of the encoder. Such a learning procedure introduces discrepant embedding representations for a same key (e.g., a code token). For instance, ASTs are composed of code tokens, which share a portion of same keys with the token vocabulary. Token names can also appear in comments. For example, more than 50\% of keys appear in both code snippets and comments vocabularies used by DeepCS and MMAN. Inconsistent embeddings for same words/tokens can lead to unsuitable matches between code snippets and comments, causing poor performance of code search (see Section~\ref{subsubsec:contribution_of_each_component}).

\begin{figure}[htbp]
  \centering
  \includegraphics[width=\linewidth]{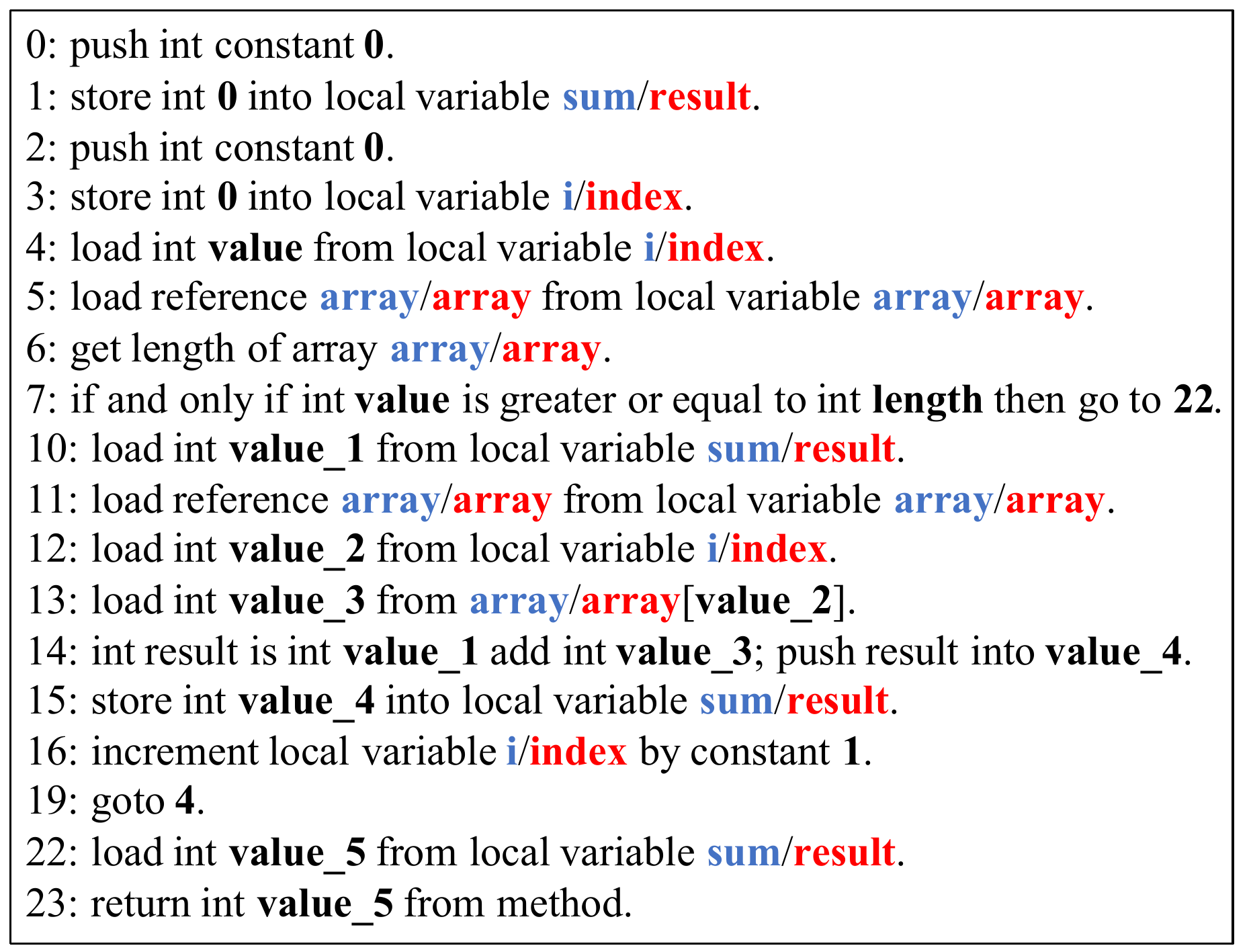}
  \caption{Code Translations of $s_a$ and $s_b$}
  \label{fig:example_of_code_translation}
\end{figure} 

\noindent
\textbf{Our solution.}
We propose a novel code search technique, called {\toolname}, that better preserves the semantics of code snippets and bridges the discrepancy between code snippets and comments. Different from existing techniques that leverage ASTs and CFGs, we directly translate code snippets into natural language sentences. Specifically, we utilize a standard program compiler and a disassembler to generate the instruction sequence of a code snippet. Such a sequence, however, lacks the context information such as local variables, data dependency, etc. We propose to simulate the execution of instructions to collect those desired contexts. A set of pre-defined translation rules are then used to translate the instruction sequence and contexts into natural language sentences. Details can be found in Section~\ref{sec:methodology}. Figure~\ref{fig:example_of_code_translation} showcases the translations of the two code snippets $s_a$ and $s_b$ by {\toolname}. The different colors denote different variable names used in $s_a$ (blue) and $s_b$ (red). The numbers/words in bold (e.g., \textbf{value} and \textbf{22}) denote the data and control dependencies among instructions. Observe that the translations of $s_a$ and $s_b$ are the same except for local variable names. The overall semantics described by the sentences in Figure~\ref{fig:example_of_code_translation} are the same. The translations are similar to those descriptions in comments, in which they share a range of words. We hence design a shared word mapping function using one single vocabulary for generating embeddings for both code snippets and comments, substantially reducing the semantic discrepancy and improving the overall performance (see results in Section~\ref{subsubsec:contribution_of_each_component}).

\section{Methodology}
\label{sec:methodology}

\begin{figure}[htbp]
  \centering
  \includegraphics[width=\linewidth]{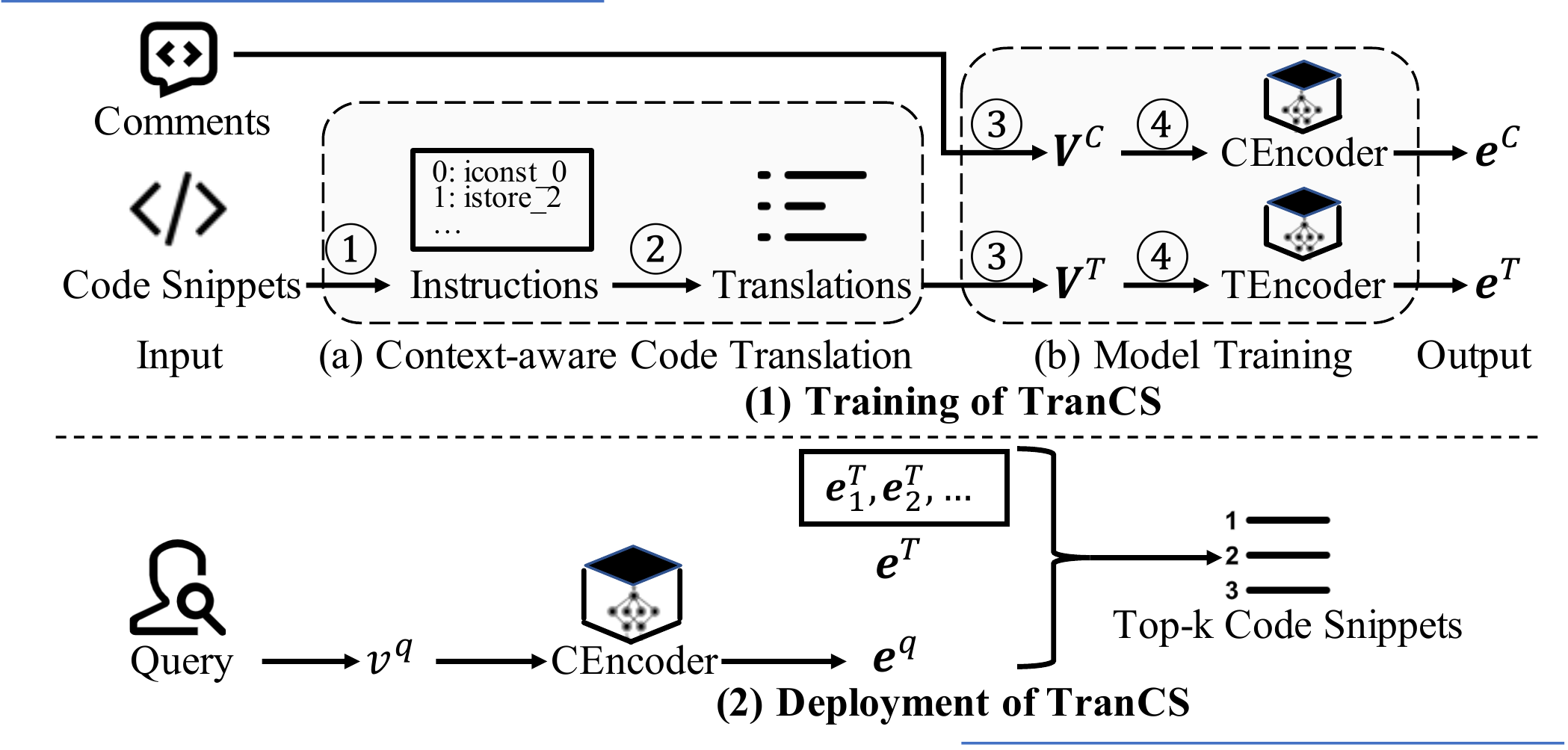}
  \caption{Framework of {\toolname}}
  \label{fig:framework_of_TranCS}
\end{figure}

\subsection{Overview}
\label{subsec:overview}
Figure~\ref{fig:framework_of_TranCS} illustrates the overview of our {\toolname}. The top part shows the training procedure of {\toolname} and the bottom part shows the usage of {\toolname} for a given query. During the training procedure of {\toolname}, two types of input data are leveraged: comments and code snippets. The comments in Figure~\ref{fig:framework_of_TranCS} are natural language descriptions that appear above the code snippet (e.g., Javadoc comments), not in the code body. These comments are input to {\toolname} in pairs with the corresponding code snippets to train CEncoder and TEncoder. For comments, {\toolname} transforms them into vector representations $\bm{V}^C$ using a shared word mapping function. For code snippets, they are different from natural language expressions such as comments. In this paper, we aim to build a homogeneous representation between comments and code snippets, which can better capture the shared semantic information of these two types. Specifically, we propose a context-aware code translation, which translates code snippets into natural language descriptions as shown in the dotted box (details are discussed in Section \ref{subsec:context-aware_code_translation}). The natural language descriptions translated from code snippets are also transformed into vector representations $\bm{V}^T$ using the same shared word mapping function. {\toolname} leverages the two vector representations $\bm{V}^C$ and $\bm{V}^T$ for building two encoders (i.e., CEncoder and TEncoder) that generate embeddings with preserved semantics for both comments and code snippets. CEncoder takes in the comment vector representations $\bm{V}^C$ and produces concise embedding representations $\bm{e}^C$ that preserves semantic information from the comments. TEncoder generates embedding representations $\bm{e}^T$ for code snippets. Details of training these two encoders are elaborated in Section \ref{subsec:mode_training}. When {\toolname} is deployed for usage, it takes in a query from the developer and passes it to CEncoder, which produces an embedding $\bm{e}^q$ for the query. {\toolname} then compares the query embedding $\bm{e}^q$ with those code embeddings $\bm{e}^T$ from the training set. A top-k selection method is leveraged for providing code snippets to the developer, which are semantically similar to the query.

\subsection{Context-aware Code Translation}
\label{subsec:context-aware_code_translation}
The goal of context-aware code translation is to translate code snippets into natural language descriptions according to the pre-defined translation rules. As shown in the dotted box of Figure \ref{fig:framework_of_TranCS}, this phase consists of two steps. In step \ding{192}, given code snippets, {\toolname} utilizes a standard compiler and disassembler to generate their instruction sequences. In step \ding{193}, {\toolname} applies the pre-defined translation rules to translate the instruction sequences into natural language descriptions. We discuss the two steps in detail in the following sections.

\subsubsection{Instruction Generation}
\label{subsubsec:instruction_generation}
\begin{figure}[htbp]
  \centering
  \includegraphics[width=\linewidth]{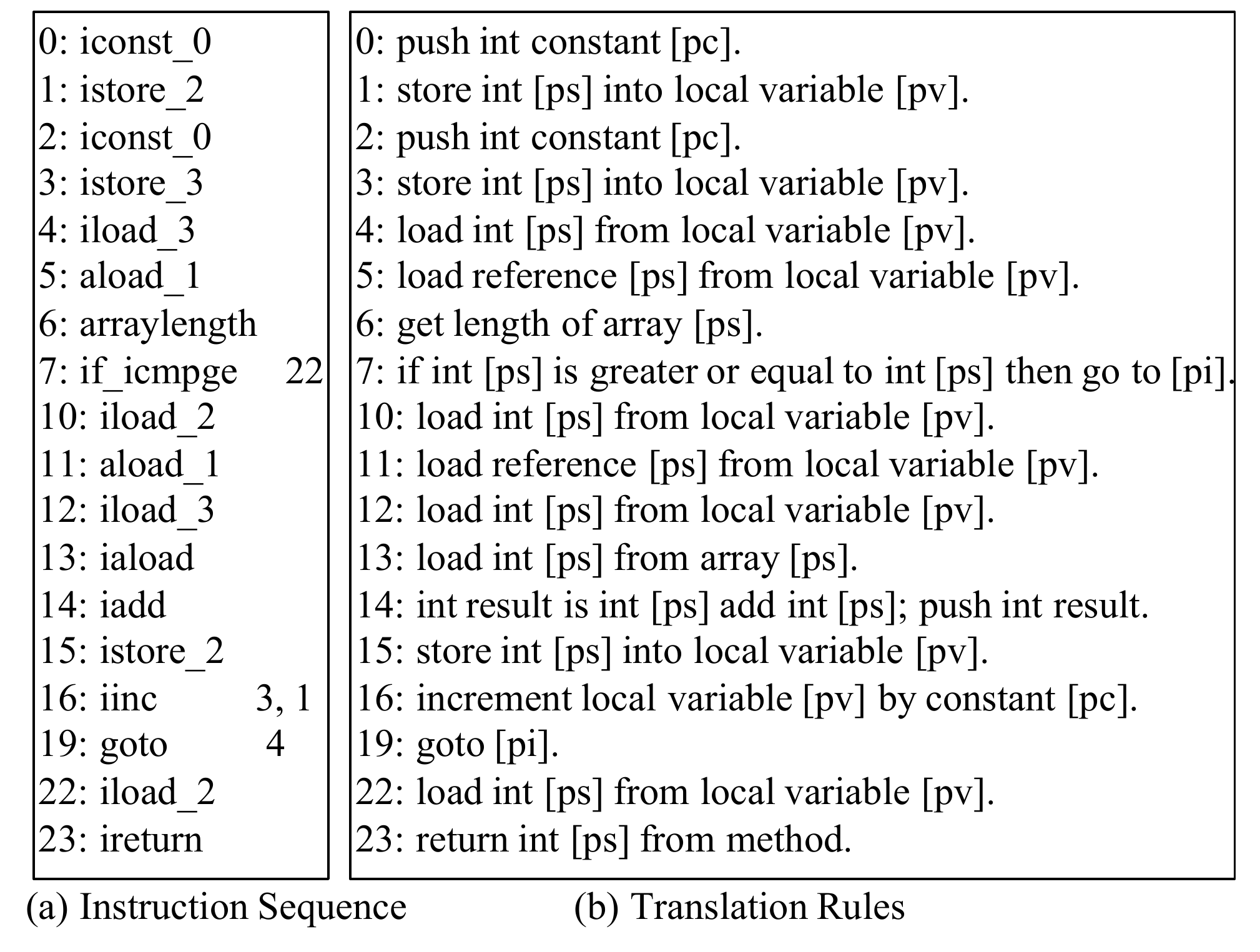}
  \caption{An Example of Instruction Sequence and Translation Rules. [pc] and [pv] indicate filling in a constant and variable, respectively. [ps] indicates filling a value popped from the operand stack, while [pi] indicates filling in an instruction index.}
  \label{fig:example_of_instruction_rules}
\end{figure}

In this step, {\toolname} takes in code snippets and produces their instruction sequences. In practice, for a given code snippet, {\toolname} first utilizes a standard program compiler and disassembler to generate the disassembly representation of the code snippet. For example, {\toolname} integrates $\mathtt{javac}$ version 1.8.0\_144 (a compiler) and $\mathtt{javap}$ version 1.8.0\_144 (a disassembler) to generate the disassembly representations for code snippets written in the Java programming language. For the code snippets that can not be compiled, the main reason is due to the lack of class/method definitions around them. We use JCoffee~\cite{2020-JCoffee} to make them compilable by adding class/method definitions around them to complement the missing pieces. Then, {\toolname} parses the disassembly representation and extracts the instruction sequence. For example, Figure \ref{fig:example_of_instruction_rules}(a) shows an instruction sequence, which is generated by inputting the code snippet shown in Figure \ref{fig:example_of_code_snippets_with_same_semantic}(a) into {\toolname}. 

In addition to the instruction sequence, {\toolname} also extracts the local variable table from the disassembly representation, which will be used in the subsequent instruction translation process. For example, Listing \ref{lst:local_variable_table} shows an example of a local variable table (LocalVariableTable) that presents the local variables involved in the code snippet in detail, and is generated along with the instruction sequence in Figure \ref{fig:example_of_instruction_rules}(a). Details about the usage of local variables are introduced in Section \ref{subsubsec:instruction_translation}.
\begin{lstlisting}[language=Java, label={lst:local_variable_table}, caption=An Example of Local Variable Table]
  LocalVariableTable:
    Start  Length  Slot  Name   Signature
        0      24     0  this   LCalArraySum;
        0      24     1 array   [I
        2      22     2   sum   I
        4      20     3     i   I
\end{lstlisting}

\subsubsection{Instruction Translation}
\label{subsubsec:instruction_translation}
In this step, {\toolname} takes in instruction sequences and produces their natural language descriptions. In this section, we first introduce the translation rules used in {\toolname}, then introduce the instruction context, and finally present how {\toolname} implements context-aware instruction translation.

\underline{\textbf{Translation Rules (TR).}}
TR used in {\toolname} is manually constructed based on the instruction specification. In practice, to construct TR, we collected all operations and descriptions of instructions from the machine instruction specification, such as Java Virtual Machine Specification~\cite{2021-JVM-Specification}. 
An operation is a short natural language description of an instruction. For example, the instruction $\mathtt{istore}$'s operation is: 
\begin{center}
\small
``store $\mathtt{int}$ into local variable.''
\end{center}
From this operation, we can know the behavior of $\mathtt{istore}$ is to store an $\mathtt{int}$ value into a local variable. A description is a long natural language description of an instruction, which details the interaction of the instruction on the local variables and operand stack. For example, $\mathtt{istore}$'s description is:
\begin{center}
\small
``The \emph{index} is an unsigned byte that must be an index into the local variable array of the current frame. The \emph{value} on the top of the operand stack must be of type $\mathtt{int}$. It is popped from the operand stack, and the value of the local variable at \emph{index} is set to value.''
\end{center}
From this description, we can know that $\mathtt{istore}$ first pops an $\mathtt{int}$ value from the operand stack and then stores the value into the \emph{index}-th position of the local variable array. If we only use the operation as the translation of the instruction, the translation will be inaccurate due to the loss of some important context. If we only use the description as the translation of instructions, the translation will be too long. However, research in the field of natural language processing (NLP) reminds us that capturing the semantics of long texts is more difficult than short texts \cite{1994-Learning-Long-term-Dependencies, 2017-Transformer}. Based on the above, we strive to make the instruction translation short and relatively accurate. Therefore, we use the operation as the basis, combing the context specified in the description, to manually collate a translation for each instruction. Such a translation delicately balances shortness and accuracy. For example, the translation we collate for the instruction $\mathtt{istore}$ as follows: 
\begin{center}
\small
``store $\mathtt{int}$ [ps] into local variable [pv].''
\end{center}
where [ps] and [pv] denote placeholders that specifies the position where the context will be filled, and details about instruction context are discussed in Section \underline{\textbf{Context-aware Instruction Translation}}. For example, Figure \ref{fig:example_of_instruction_rules}(b) shows the result of {\toolname} using TR to translate the instruction sequence in Figure \ref{fig:example_of_instruction_rules}(a).

\begin{figure}[htbp]
  \centering
  \includegraphics[width=\linewidth]{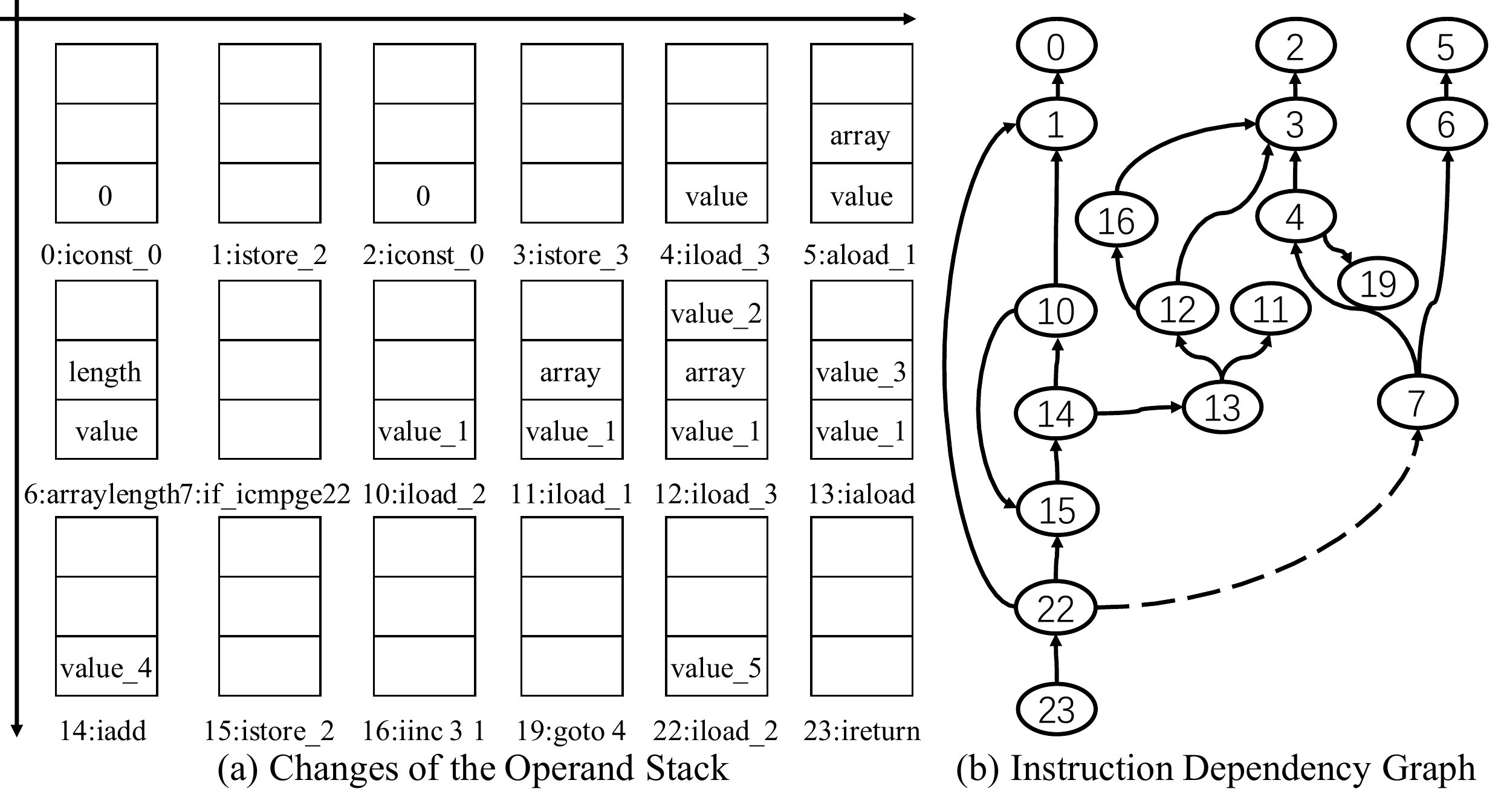}
  \caption{An Example of the Changes of the Operand Stack and Instruction Dependency Graph}
  \label{fig:example_of_OS_IDG}
\end{figure}

\underline{\textbf{Instruction Context.}}
The context of an instruction consists of constants, local variables, and data and control dependencies with other instructions. Constants and local variables are directly determined by operands. As shown in Figure \ref{fig:example_of_instruction_rules}(a), an opcode is followed by zero or more operands. An operand can be a constant, or an index of a local variable, or an index of an instruction. For example, in Figure \ref{fig:example_of_instruction_rules}(a), the operand 0 following the opcode $\mathtt{iconst}$ represents a constant, while the operand 2 following the opcode $\mathtt{istore}$ represents the index of the local variable $sum$ shown in Listing \ref{lst:local_variable_table}; Control dependencies between instructions are explicitly passed through the indices of the instruction. The indices are also directly specified by operands. For example, the operand 22 following the opcode $\mathtt{if\_icmpge}$ represents the index of the instruction $\mathtt{iload\_2}$ at line 22 in Figure \ref{fig:example_of_instruction_rules}(a). Data dependencies between instructions are implicitly passed through the operand stack. As described in Section \underline{\textbf{Translation Rules (TR)}}, with the guidance of the description, we can know how each instruction interacts with the operand stack, such as popping or pushing data. If the instruction $i_a$ pops (i.e., uses) the data that is pushed onto the operand stack by the instruction $i_b$, then we say that $i_a$ is data dependent on $i_b$. For example, Figure \ref{fig:example_of_OS_IDG}(a) shows the changes of the operand stack as the opcode sequence in Figure \ref{fig:example_of_instruction_rules} interacts with the operand stack. The values in the operand stack are the carriers that reflect data dependencies between instructions. Figure \ref{fig:example_of_OS_IDG}(b) shows the data and control dependencies between the instructions in Figure \ref{fig:example_of_instruction_rules}(a). In this figure, nodes represent instructions; the labels of nodes are instructions' indices; the solid and dashed edges represent data and control dependencies, respectively. 

\underline{\textbf{Context-aware Instruction Translation.}} 
The basic idea of context-aware instruction translation is to simulate the execution of instructions by statically traversing the instruction sequence from top to down. In the traversal process, we collect the context of each instruction, which will be used to update the TR-based translations of the current or other related instructions. 

In the actual execution of instructions, a frame is created when the corresponding code snippet is invoked \cite{2021-JVM-Specification-Frame}. A frame contains a local variable array and a last-in-first-out stack (i.e., operand stack). The sizes of the local variable array and the operand stack are determined at compile-time. The local variable array stores all local variables used in the instructions. For example, the local variables shown in Listing \ref{lst:local_variable_table} are used in the instruction sequence shown in Figure \ref{fig:example_of_instruction_rules}(a). The indices of the local variable array corresponds to that in LocalVariableTable shown in Listing \ref{lst:local_variable_table}, where the `Slot' column presents indices of the local variables. The names and indices of local variables are determined at compile-time, but their values are dynamically updated with the execution of the instructions. The values in the operand stack are also dynamically updated with the execution of the instructions. As mentioned earlier, the context of an instruction includes constants, local variables, data and control dependencies. Among them, constants, local variables and control dependencies are closely related to instructions' operands. They can be easily determined by the operands (for constants) or by retrieving the instruction sequence (for control dependencies) using the index specified by the operand. However, determining the values of local variables is a challenging task because they are dynamically updated with the execution of the instruction. Analogously, the determination of data dependencies is a challenging task because they are implicitly passed through the operand stack. The values in operand stack are also dynamically updated with the execution of the instruction. Therefore, we need to know in advance how the instruction interacts with the local variable array (e.g., setting value) or the operand stack (e.g., popping or pushing data). In practice, we obtain such information from the description of each instruction. The description of each instruction has been introduced when we introduced the translation rules earlier. With the guidance of the description, we divide the instructions into the following four categories according to whether they interact with the local variable array or the operand stack.

Category 1, expressed as $\mathbb{I}^{S}$. 
In $\mathbb{I}^{S}$, the instruction only interacts with the operand stack. 
$\mathbb{I}^{S}$ can be subdivided into the following three types:
\begin{description}
  \item[$\mathbb{I}^{PU}$.] In this type, the interaction is to push the operand onto the operand stack.
  \item[$\mathbb{I}^{PO}$.] In this type, the interaction is to pop values from the operand stack.
  \item[$\mathbb{I}^{POU}$.] In this type, the interaction is composed of popping values from the operand stack, performing the operation, and pushing the result of the operation to the operand stack.
\end{description}

Category 2, $\mathbb{I}^{V}$. 
In $\mathbb{I}^{V}$, the instruction only interacts with the local variable array.
The interaction is to load the value from the local variable array, or store the new value into it. This type of instruction does not interact with the operand stack. For example, the instruction $\mathtt{iinc \;3, 1}$ only interacts with the local variable specified by the first operand, not with the operand stack.

Category 3, $\mathbb{I}^{SV}$.
In $\mathbb{I}^{SV}$, the instruction interacts with the operand stack as well as the local variable array. For example, the instruction $\mathtt{istore\_2}$ first loads the integer value from the operand stack, and then stores the value into a local variable. 

Category 4, $\mathbb{I}^{O}$. 
In $\mathbb{I}^{O}$, the instruction neither interacts with the operand stack nor with the local variable array, such as the instruction $\mathtt{goto}$ and $\mathtt{nop}$. Table \ref{tab:category_of_opcode} shows the categories of instructions. 

\begin{table*}[htbp]
\small
  \begin{center}
  \renewcommand{\arraystretch}{1.2}
  \caption{The Category of Instructions}.
  \begin{tabular}{|c|c|c|c|c|c|c|}
    \hline
    \textbf{Category} &\multicolumn{3}{c|}{$\mathbb{I}^{S}$} & \multirow{2}*{$\mathbb{I}^{V}$} & \multirow{2}*{$\mathbb{I}^{SV}$} & \multirow{2}*{$\mathbb{I}^{O}$} \\
     \cline{1-4}
     \textbf{Type} & $\mathbb{I}^{PU}$ & $\mathbb{I}^{PO}$ & $\mathbb{I}^{POU}$ & & & \\
     \hline
     \textbf{Instructions}
      & \tabincell{c}{aconst\_null, \\ anewarray, \\ iconst, fconst, \\ bipush, \\ dconst\_<d>, \\ fconst\_<f>, \\ iconst\_<i>, \\ jsr, jsr\_w, \\ lconst\_<l>, \\ ldc, ldc\_w, \\ ldc2\_w, new, \\sipush} 
      & \tabincell{c}{areturn, if\_icmpge, \\ ireturn, athrow, \\ dreturn, freturn, \\ if\_acmp<cond>, \\ if\_icmp<cond>, \\ if<cond>, ifnonnull, \\ ifnull, invokedynamic, \\ invokeinterface, \\ invokespecial, invokestatic, \\ invokevirtual, ireturn, \\ ishl, ishr, lookupswitch, \\ lreturn, monitorexit, pop, \\ pop2, putfield, \\ putstatic, tableswitch}  
      & \tabincell{c}{aaload, arraylength, baload, \\ caload, d2f, d2i, d2l, dadd, \\ daload, dcmp<op>, ddiv, dmul, \\ dneg, drem, dsub, dup, dup\_x1, \\ dup\_x2, dup2, dup2\_x1, dup2\_x2, \\ f2d, f2i, f2l, fadd, faload, fcmp<op>, \\ fdiv, fmul, fneg, frem, fsub, getfield, \\ getstatic, i2b, i2c, i2d, i2f, i2l, i2s, \\ iadd, iaload, iand, idiv, imul, ineg, \\ instanceof, ior, irem, isub, iushr, \\ ixor, l2d, l2f, l2i, ladd, laload, land, \\ lcmp, ldiv, lmul, lneg, lor, lrem, \\ lshl, lshr, lsub, lushr, multianewarray, \\ lxor, newarray, saload, swap}  
      & \tabincell{c}{iinc, \\ wide} 
      & \tabincell{c}{aastore, aload, \\aload\_<n>, astore \\ astore\_<n>, \\ bastore, castore, dastore, \\ dload, dload\_<n>, \\ dstore, dstore\_<n>, \\ fastore, fload, \\ fload\_<n>, fstore, \\ fstore\_<n>, iastore, \\ iload, iload\_<n>, \\ istore, istore\_<n>, \\ lastore, lload, \\ lload\_<n>, lstore, \\ lstore\_<n>, sastore } 
      & \tabincell{c}{goto, \\ checkcast, \\goto\_w, \\ nop, \\ ret, \\ return}\\
     \hline
    \end{tabular}
  \label{tab:category_of_opcode}
  \end{center}
\end{table*}

\begin{algorithm}[htbp]
\small
  \caption{Context-aware Instruction Translation}
  \label{alg:instruction_translator}
  \begin{algorithmic}[1]
    \REQUIRE An instruction sequence, $I$; Translation Rules, $TR$; \\
                    A local variable array, $V$; 
                    The depth of the operand stack, $d$.
    \ENSURE Instruction Translation, $T$;
    \STATE $S \leftarrow $ initialize an empty operand stack with a depth of $d$.
    \FOR{\textbf{each} \rm{$i$ in $I$}}
      \STATE $t \leftarrow $ generate the TR-based translation of $i$ based on $TR$;
      \STATE $operands \leftarrow $ extract the operands from $i$;
      \IF {$i \in \mathbb{I}^{PU}$}
        \STATE $S \leftarrow$ push $operands$ onto $S$;
        \STATE $t \leftarrow $ replace [pc] in $t$ with $operands$;
      \ENDIF
      \IF {$i \in \mathbb{I}^{PO}$}
         \STATE $values \leftarrow$ pop values from $S$ by $operands$;
         \STATE $t \leftarrow $ replace [ps] in $t$ with $values$;
      \ENDIF
      \IF {$i \in \mathbb{I}^{POU}$}
         \STATE $values \leftarrow $ pop values from $S$ by $operands$;
         \STATE $t \leftarrow $ replace [ps] in $t$ with $values$;
         \STATE $new\_value \leftarrow $ do operation;
         \STATE  $S \leftarrow$ push $new\_value$ onto $S$;
      \ENDIF
      \IF {$i \in \mathbb{I}^{V}$}
        \STATE $variable \leftarrow $ get variable from $V$ by $operands$;
      	\STATE $t \leftarrow $ replace [pv] in $t$ with $variable$;
      \ENDIF
      \IF {$i \in \mathbb{I}^{SV}$}
        \STATE $values \leftarrow $ pop values from $S$ by $operands$;
        \STATE $t \leftarrow $ replace [ps] in $t$ with $values$;
        \STATE $variable \leftarrow $ get variable from $V$ by $operands$;
         \STATE $t \leftarrow $ replace [pv] in $t$ with $variable$;
      \ENDIF
      \IF {$t$ contains [pi]}
         \STATE $t \leftarrow $ replace [pi] in $t$ with $operands$;
      \ENDIF
      \STATE $T \leftarrow T \cup \{t\}$
    \ENDFOR
    \STATE \text{output} $T$;
  \end{algorithmic}
\end{algorithm}

Based on the above classification, {\toolname} uses Algorithm~\ref{alg:instruction_translator} to perform context-aware instruction translation. {\toolname} takes an instruction sequence ($I$), translation rules ($TR$), a local variable array ($V$), and the depth of the operand stack ($d$) as inputs. $TR$, $V$ and $d$ have been introduced earlier. {\toolname} first initializes an stack with a depth of $d$ to store intermediate results produced during traversing $I$ (line 1). {\toolname} then traverses $I$ from top to down (lines 2 -- 33). 
For each $i \in I$, {\toolname} first generates its translation $t$ based on $TR$ (line 3). 
Then, {\toolname} extracts the operands from $i$ (line 4). 
The operands are used to update $S$ and $t$ in subsequent processes. 
{\toolname} determines $i$'s category according to the pre-defined categories shown in Table \ref{tab:category_of_opcode}.
According to $i$'s category, {\toolname} uses different processes to update $S$ and $t$ (line 5 -- 31). 
For example, Figure \ref{fig:example_of_OS_IDG}(a) shows an example of the changes of the operand stack when {\toolname} traverses the instruction sequence shown in Figure \ref{fig:example_of_instruction_rules}(a) from top to down. 
After traversing all the instructions in $I$, the algorithm finishes and outputs $I$'s translations $T$.  
For example, Figure~\ref{fig:example_of_code_translation} shows the translation generated by {\toolname} for the instruction sequence shown in Figure~\ref{fig:example_of_instruction_rules}(a).

\subsection{Model Training}
\label{subsec:mode_training}
The goal of model training is to train two encoders, which will be deployed to support code search service. This phase consists of two steps as shown in Figure~\ref{fig:framework_of_TranCS}. In step \ding{194}, given translations and comments, {\toolname} transforms them into vector representations $\bm{V}^C$ and $\bm{V}^T$ using a shared word mapping function. In step \ding{195}, {\toolname} leverages $\bm{V}^C$ and $\bm{V}^T$ to train CEncoder and TEncoder. 

\subsubsection{Shared Word Mapping}
\label{subsubsec:shared_word_mapping}
In {\toolname}, both translations and comments are natural language sentences. Sentence embedding is generated based on word embedding~\cite{2013-Word2vec, 2015-Deep-Sentence-Embedding}. Word embedding techniques can map words into fixed-length vectors (i.e., embeddings) so that similar words are close to each other in the vector space~\cite{2013-Estimation-Word-Representations, 2013-Word2vec}. 

A word embedding technique can be considered a word mapping function $\psi$, which can map a word $w_i$ into a vector representation $\bm{w}_i$, i.e., $\bm{w_i} = \psi(w_i)$. As aforementioned, both translations and comments are natural language sentences, so we design a shared word mapping function. To implement such a $\psi$, we build a shared vocabulary that includes top-$n$ frequently appeared words in translations and comments. We further transform the vector representations of the words into an embedding matrix $E \in \mathbb{R}^{n \times m}$, where $n$ is the size of the vocabulary, $m$ is the dimension of word embedding. The embedding matrix $E = (\psi(\bm{w}_1), ..., \psi(\bm{w}_i))^T$ is initialized randomly and learned in the training process along with the two encoders. Based on this embedding matrix, {\toolname} can transforms translations and comments into the vector representations $\bm{V}^C$ and $\bm{V}^T$. A simple way of sentence vector representations is to view it as a bag of words and add up all its word vector representations~\cite{2014-Distributed-Representations-of-Sentences}. 

\subsubsection{Encoder Training}
\label{subsubsec:encoder_training}
In this section, we first introduce the architecture of CEncoder and TEncoder, then present how to jointly train the two encoders.

\textbf{Encoder Architecture.}
As described in Section\ref{subsubsec:shared_word_mapping}, in {\toolname} both translations and comments are natural language sentences. Therefore, we can use the same sequence embedding network to design comment encoder (CEncoder) and translation encoder (TEncoder) instead of designing different embedding networks for them as the previous DL-based CS techniques, such as DeepCS~\cite{2018-DeepCodeSearch} and MMAN~\cite{2019-Multi-modal-Attention-for-Code-Rerieval}. In practice, {\toolname} applies the LSTM architecture to design CEncoder and TEncoder. Consider a translation/comment sentence $s=w_1, \cdots, w_{N^{s}}$ comprising a sequence of $N^{s}$ words, {\toolname} first uses the shared word mapping function to produce vector representations $\bm{v}^{s}$. Then, {\toolname} passes $\bm{v}^{s}$ to the encoder (i.e., CEncoder or TEncoder) that generates embeddings $\bm{e}^{s}$. The hidden state $\bm{h}_i^{s}$ of the $i$-th word in $s$ is calculated as follows:
\begin{equation}
  \bm{h}_{i}^s = LSTM(\bm{h}_{i-1}^s, \bm{w}_i)
  \label{equ:lstm}
\end{equation}
where $\bm{w}_i$ represents the vector of the word $w_i$ and comes from the embedding matrix $E$. 

In addition, {\toolname} uses attention mechanism proposed by Bahdanau et al.~\cite {2015-Attention-Neural-Machine-Translation} to alleviate the long-dependency problem in long text sequences~\cite{1994-Learning-Long-term-Dependencies}. The attention weight for each word $w_i$ is calculated as follows:
\begin{equation}
  \alpha^s_i = \frac{exp(f(\bm{h}^s_i) \cdot \bm{u}^s)}{\sum_{j=1}^{N^s}{exp(f(\bm{h}^s_j) \cdot \bm{u}^s)}}
  \label{equ:attention}
\end{equation}
where $f(\cdot)$ denotes a linear layer; 
$\bm{u}^s$ denotes the context vector which is a high level representation of all words in $s$; 
and $\cdot$ denotes the inner project of $\bm{h}^s_i$ and $\bm{u}^s$. 
The context vector $\bm{u}^s$ is randomly initialized and jointly learned during training. 
Then, $s$'s final embedding representation $\bm{e}^s$ can be calculated as follows: 
\begin{equation}
  \bm{e}^s = \sum_{j=1}^{N^s}{\alpha^s_i \cdot \bm{h}^s_i}
  \label{equ:e_com}
\end{equation}

\textbf{Joint Training.}
Now we present how to jointly train the two encoders (i.e., CEncoder and TEncoder) of {\toolname} to transform both translations and comments into a unified vector space with a similarity coordination. We follow a widely adopted assumption that if a translation and a comment have similar semantics, their embedding representations should be close to each other~\cite{2018-DeepCodeSearch, 2019-Multi-modal-Attention-for-Code-Rerieval, 2020-Code-Search-with-Co-Attentive-Representation}. In other words, given a code snippet $s$ whose translation is $t$ and a comment $c$, we want it to predict a high similarity between $t$ and $c$ if $c$ is a correct comment of $s$, and a little similarity otherwise.

In practice, we first translate all code snippets into translations. Then, we construct each training instance as a triple $\langle t, c^+, c^- \rangle$: for each translation $t$ there is a positive comment $c^+$ (a ground-truth comment of $s$) and a negative comment $c^-$ (an incorrect comment of $s$). The incorrect comment $c^-$ is selected randomly from the pool of all correct comments. 
When trained on the set of $\langle t, c^+, c^- \rangle$ triples, {\toolname} predicts the cosine similarities of both $\langle t, c^+\rangle$ and $\langle t, c^- \rangle$ pairs and minimizes the ranking loss \cite{2011-NLP-Almost-from-Scratch, 2013-DeViSE}:
\begin{equation}
  \mathcal{L}(\theta) = \sum_{\langle \bm{t}, \bm{c^{+}}, \bm{c^-}\rangle \in G}{max(0, \beta - cos(\bm{t}, \bm{c^{+}}) + cos(\bm{t}, \bm{c^-}))}
  \label{equ:l_theta}
\end{equation}
where $\theta$ denotes the model parameters; 
$G$ denotes the training dataset; 
$\beta$ is a small and fixed margin constraint;
$\bm{t}$, $\bm{c^{+}}$ and $\bm{c^-}$ are the embedded vectors of $t$, $c^+$ and $c^-$, respectively. 
Intuitively, the ranking loss encourages the cosine similarity between a translation and its correct comment to go up, and the cosine similarities between a translation and incorrect comments to go down.

\subsection{Deployment of {\toolname}}
\label{subsec:model_optimization}
After the two encoders (i.e., CEncoder and TEncoder) are trained, we can deploy {\toolname} online for code search service. Figure~\ref{fig:framework_of_TranCS}(2) shows the deployment of {\toolname}. For a search query $q$ given by the developer, {\toolname} first uses the shared word mapping function to transform it into vector representation $\bm{v}^q$. {\toolname} further passes $\bm{v}^q$ into CEncoder to generate the embedding $\bm{e}^q$. Then, {\toolname} measures the similarity between $\bm{e}^q$ and each $\bm{e}^t \in \bm{e}^T$. 
The similarity is calculated as follows: 
\begin{equation}
  sim(q, t) = cos(\bm{e}^q, \bm{e}^t) = \frac{\bm{e}^q \cdot \bm{e}^t}{\left \| \bm{e}^q \right \| \left \| \bm{e}^t \right \|}
  \label{equ:cosine_similarity}
\end{equation}
{\toolname} ranks all $\bm{T}$ by their similarities with $q$. The higher the similarity, the higher relevance of the code snippet to $q$. Finally, {\toolname} outputs the code snippets corresponding to the top-$k$ translations to the developer.

\section{Evaluation And Analysis}
\label{sec:evaluation}
We conduct experiments to answer the following questions:
\begin{description}
	\item[RQ1.] What is the effectiveness of {\toolname} when compared with state-of-the-art techniques?
	\item[RQ2.] What is the contribution of key components in {\toolname}, i.e., context-aware code translation and shared word mapping?
	\item[RQ3.] What is the robustness of {\toolname} when varying the query length and code length?
\end{description}

\subsection{Experimental Setup}
\label{subsubsec:experimental_setup}

\subsubsection{Dataset}
\label{subsubsec:dataset}
We evaluate the performance of our {\toolname} on a corpus of Java code snippets, collected from the public CodeSearchNet corpus~\cite{2019-CodeSearchNet-Dataset}. Actually, we have considered the dataset released by baselines (i.e., DeepCS~\cite{2018-DeepCodeSearch} and MMAN~\cite{2019-Multi-modal-Attention-for-Code-Rerieval}). However, the dataset of DeepCS only contains the cleaned Java code snippets without the raw data, unable to generate the CFG for MMAN. And the dataset of MMAN is not publicly accessible. 

\begin{figure}[!t]
\graphicspath{{pictures/}}
\centering
    \subfigure[Code Snippets on Training Set]
    {
        \includegraphics[width=0.475\linewidth]{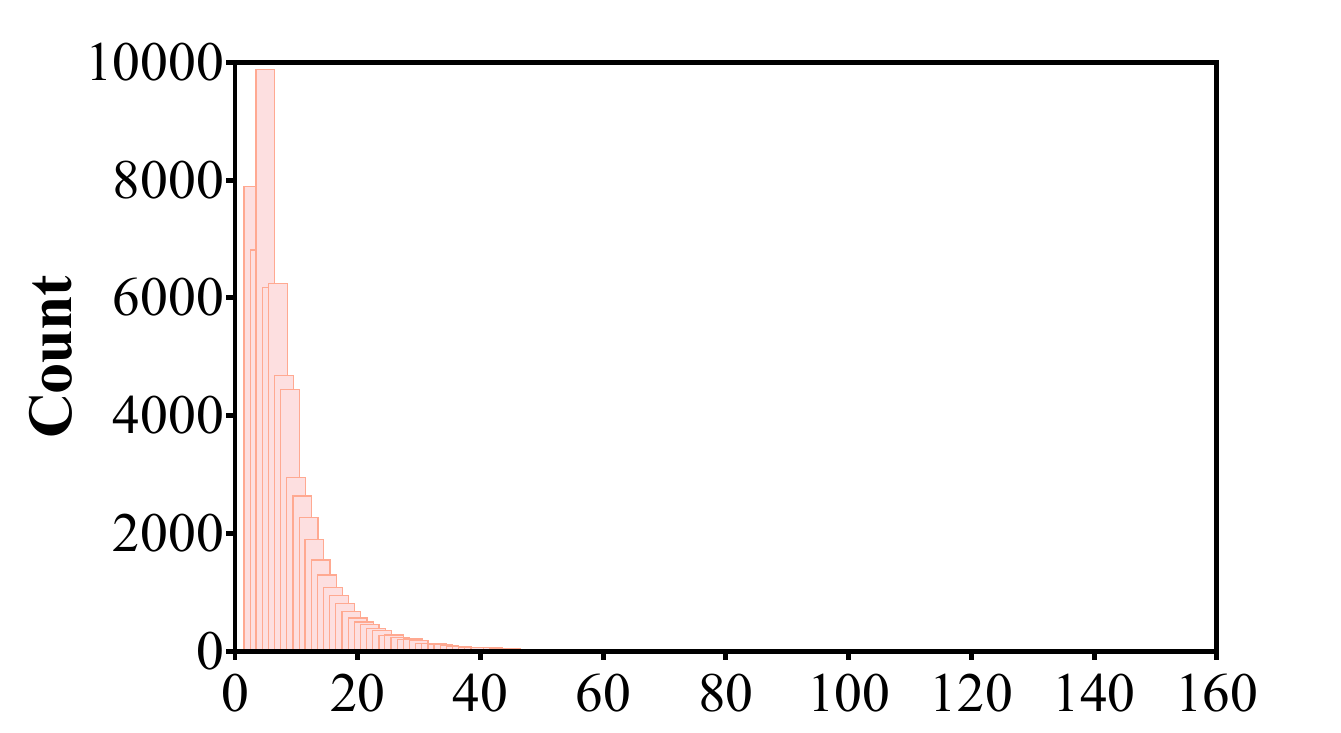}
        \label{fig:distribution_of_code_length_train_set}
    }
    \subfigure[Comments on Training Set]
    {
        \includegraphics[width=0.475\linewidth]{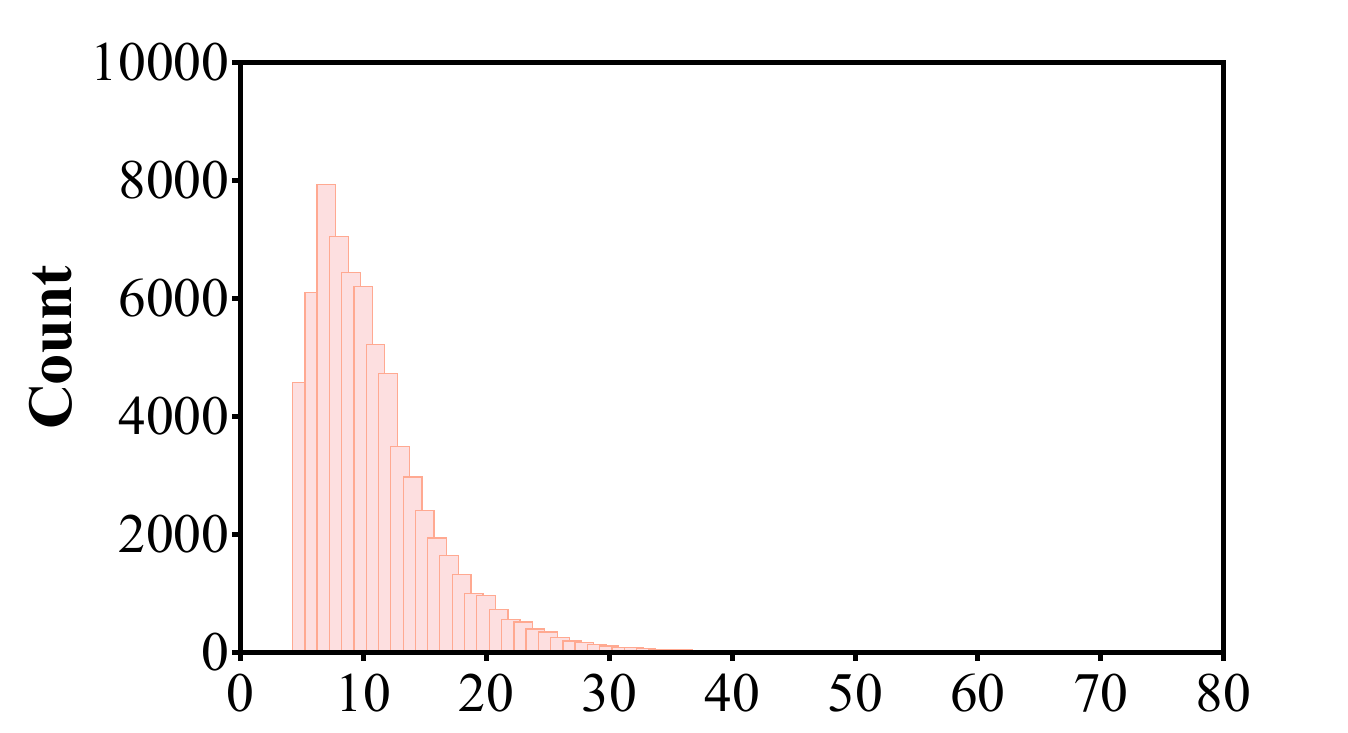}
        \label{fig:distribution_of_comment_length_train_set}
    }
        \subfigure[Code Snippets on Testing Set]
    {
        \includegraphics[width=0.475\linewidth]{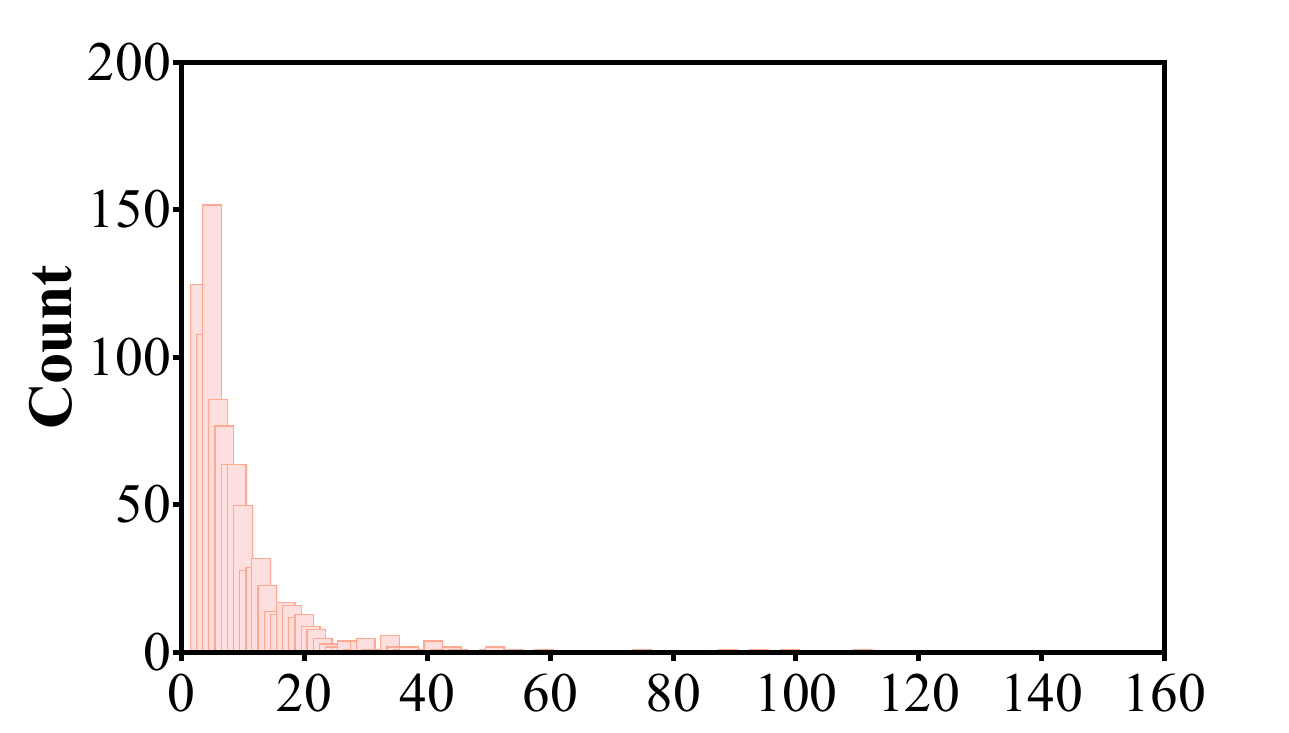}
        \label{fig:distribution_of_code_length_test_set}
    }
    \subfigure[Comments on Testing Set]
    {
        \includegraphics[width=0.475\linewidth]{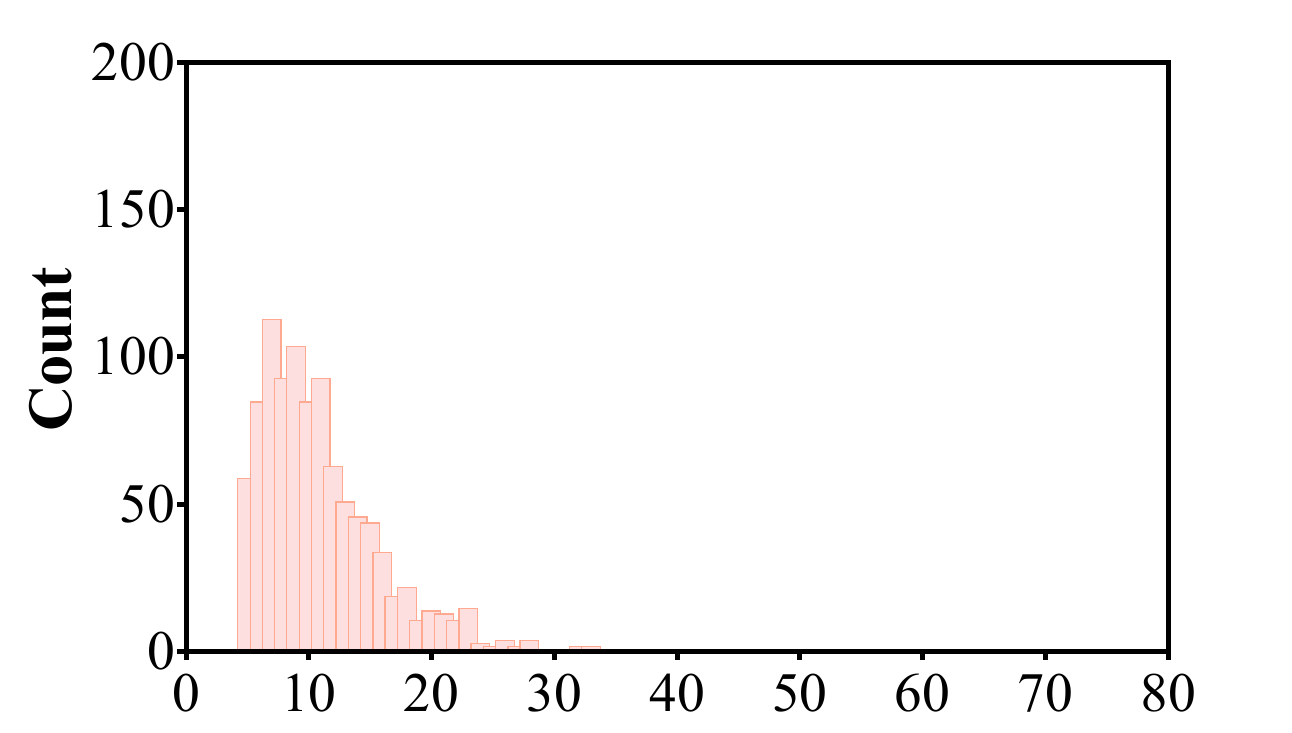}
        \label{fig:distribution_of_comment_length_test_set}
    }
    \caption{Length Distributions}
    \label{fig:distribution_of_length}
\end{figure}

We randomly shuffle the dataset and split it into two parts, i.e., 69,324 samples for training and 1,000 samples for testing. It is worth mentioning a difference between our data processing and the one in~\cite{2018-DeepCodeSearch}. 
In~\cite{2018-DeepCodeSearch}, the proposed approach is verified on another isolated dataset to avoid the bias. Since the evaluation dataset does not have the ground truth, they manually labelled the searched results. 
As possible subjective bias exists in manual evaluation~\cite{2019-DL-Met-CodeSearch, 2019-Multi-modal-Attention-for-Code-Rerieval}, in this paper, we also adopt the automatic evaluation.
Figure \ref{fig:distribution_of_length}(a) and (b) show the length distributions of code snippets and comments on the training set. For a code snippet, its length refers to the number of lines of the code snippet. For a comment, its length refers to the number of words in the comment. From Figure~\ref{fig:distribution_of_length}(a), we can observe that the lines of most code snippets are located between 20 to 40. This was also observed in the quote in~\cite{2009-Clean-Code} ``Functions should hardly ever be 20 lines long''. From Figure \ref{fig:distribution_of_comment_length_train_set}, it is noticed that almost all comments are less than 20 in length.  
This also confirms the challenge of capturing the correlation between short text with its corresponding code snippet. 
Figure \ref{fig:distribution_of_length}(c) and (d) show the length distributions of code snippets and comments on testing data. 
We can observe that, despite shuffling randomly, the distributions of data sizes (i.e., lengths) in the two data sets are consistent, so we can conclude that the testing set is representative.

\subsubsection{Evaluation Metrics}
\label{subsubsec:evaluation_metrics}
In the evaluation, we consider the comment of the code snippet as the query, and the code snippet itself as the ground-truth result of code search, which is similar to~\cite{2019-Multi-modal-Attention-for-Code-Rerieval, 2020-Code-Search-with-Co-Attentive-Representation, 2020-Multi-Perspective-Architecture-for-CS} but different from~\cite{2018-DeepCodeSearch, 2019-DL-Met-CodeSearch}. 
During the testing time, we treat each comment in the 1,000 testing samples as a query, the code snippet corresponding to the query as the correct result, and the other 999 code snippets as distractor results. We adopt two automatic evaluation metrics that are widely used in code search studies~\cite{2018-DeepCodeSearch, 2019-Multi-modal-Attention-for-Code-Rerieval, 2019-DL-Met-CodeSearch, 2020-Code-Search-with-Co-Attentive-Representation, 2020-Multi-Perspective-Architecture-for-CS} to measure the performance of {\toolname}, i.e., success rate at $k$ ($\mathtt{SuccessRate@k}$) and mean reciprocal rank ($\mathtt{MRR}$).

$\mathtt{SuccessRate@k}$ measures the percentage of queries for which the correct result exists in the top $k$ ranked results~\cite{2017-Re-evaluating-Automatic-Metrics, 2019-Multi-modal-Attention-for-Code-Rerieval}, which is computed as follows:
\begin{equation}
	\mathtt{SuccessRate@k} = \frac{1}{|Q|}\sum_{i = 1}^{|Q|}{\delta(FRank_{Q_i} \leq k)}
	\label{equ:SuccessRate}
\end{equation}
where $Q$ denotes a set of queries and $|Q|$ is the size of $Q$; $\delta(\cdot)$ denotes a function which returns 1 if the input is true and returns 0 otherwise; $FRank_{Q_i}$ refers to the rank position of the correct result for the $i$-th query in $Q$. $\mathtt{SuccessRate@k}$ is important because a better CS technique should allow developers to discover the expected code snippets by inspecting fewer returned results.  
The higher the $\mathtt{SuccessRate@k}$ value, the better the code search performance.

$\mathtt{MRR}$ is the average of the reciprocal ranks of results of a set of queries $Q$~\cite{2019-Multi-modal-Attention-for-Code-Rerieval, 2020-Code-Search-with-Co-Attentive-Representation}. The reciprocal rank of a query is the inverse of the rank of the correct result. 
$\mathtt{MRR}$ is computed as follows:
\begin{equation}
	\mathtt{MRR} = \frac{1}{|Q|}\sum_{i = 1}^{|Q|}{\frac{1}{FRank_{Q_i}}}
	\label{equ:MRR}
\end{equation} 
The higher the $\mathtt{MRR}$ value, the better the code search performance.

Meanwhile, as developers prefer to find the expected code snippets with short inspection, we only test $\mathtt{SuccessRate@k}$ and $\mathtt{MRR}$ on the top-10 (that is, the maximum value of $k$ is 10) ranked list following DeepCS~\cite{2018-DeepCodeSearch} and MMAN~\cite{2019-Multi-modal-Attention-for-Code-Rerieval}. 
In other words, when the rank of $Q_i$ is out of 10, then $1/FRank_{Q_i}$ is set to 0.

\subsubsection{Baselines.}
\label{subsubsec:baselines}
In this paper, we compare the following baselines:
\begin{itemize}
  \item \textbf{DeepCS}~\cite{2018-DeepCodeSearch}. 
  DeepCS is one of the representative DL-based CS techniques. 
  DeepCS uses two kinds of model architecture to design the code encoder to embed three aspects of the code snippet, i.e., two RNNs for method names and API sequences, and a multi-layer perceptron (MLP) for tokens. 
  Its query encoder also uses RNN architecture. 
  \item \textbf{MMAN}~\cite{2019-Multi-modal-Attention-for-Code-Rerieval}. 
  MMAN is one of the state-of-the-art DL-based CS techniques. 
  MMAN uses multiple kinds of model architectures to design the code encoder to embed multiple aspects of the code snippet, i.e., one LSTM for Token, a Tree-LSTM for AST, and a GGNN for CFG. 
  Its query encode uses LSTM architecture.
\end{itemize}

\subsubsection{Implementation Details}
\label{subsubsec:implementation_details}
To train our model, we first shuffle the training data and set the mini-batch size to 32. 
The size of the vocabulary is 15,000. 
For each batch, the code snippet is padded with a special token $\langle PAD \rangle$ to the maximum length. 
We set the word embedding size to 512. 
For LSTM unit, we set the hidden size to 512. 
The margin $\beta$ is set to 0.6. 
We update the parameters via AdamW optimizer~\cite{2015-Adam} with the learning rate 0.0003. 
To prevent over-fitting, we use dropout with 0.1.
In {\toolname}, the comment and the code snippet share the same embedding weights. 
All models are implemented using the PyTorch 1.7.1 framework with Python 3.8. 
All experiments are conducted on a server equipped with one Nvidia Tesla V100 GPU with 31 GB memory, running on Centos 7.7. 
All the models in this paper are trained for 200 epochs, and we select the best model based on the lowest validation loss.

\subsection{Evaluation Results}
\label{subsec:evaluation_results}
In this section, we present and analyze the experimental results to answer the research questions. 

\subsubsection{\textbf{RQ1:} Effectiveness of {\toolname}}
\label{subsubsec:effectiveness_of_TranCS}
Table \ref{tab:compare_with_baselines} shows the overall performance of {\toolname} and two baselines, measured in terms of $\mathtt{SuccessRate@k}$ and $\mathtt{MRR}$. 
The columns $\mathtt{SR@1}$, $\mathtt{SR@5}$ and $\mathtt{SR@10}$ show the results of the average $\mathtt{SuccessRate@k}$ over all queries when $k$ is 1, 5 and 10, respectively. 
The column $\mathtt{MRR}$ shows the $\mathtt{MRR}$ values of the three techniques. 
From this table, we can observe that for $\mathtt{SR@k}$, the improvements of {\toolname} to DeepCS are 102.90\%, 45.80\% and 32.48\% when $k$ is 1, 5, and 10, respectively. 
The improvements to MMAN are 67.16\%, 35.94\%, and 25.42\%, respectively. 
For $\mathtt{MRR}$, the improvements {\toolname} to DeepCS and MMAN are 66.50\% and 49.31\%, respectively. 
We can draw the conclusion that under all experimental settings, our {\toolname} consistently achieves higher performance in terms of both two metrics, which indicates better code search performance. 

\begin{table}[htbp]
  \renewcommand{\arraystretch}{1.0}
  \caption{Overall Performance of {\toolname} and Baselines}
  \label{tab:compare_with_baselines}
  \centering
  \begin{tabular}{c|ccc|c}
    \toprule
     Tech & $\mathtt{SR@1}$ & $\mathtt{SR@5}$ & $\mathtt{SR@10}$ & $\mathtt{MRR}$\\
    \midrule
    DeepCS & 0.276 & 0.524 & 0.622 & 0.391 \\
    MMAN & 0.335 & 0.562 & 0.657  & 0.436 \\
    \midrule
    {\toolname} & \textbf{0.560} & \textbf{0.764} & \textbf{0.824}  & \textbf{0.651} \\
    \bottomrule
 \end{tabular}
\end{table}

The CodeSearchNet corpus also provides 99 realistic natural languages queries and expert annotations for likely results. Each query/result pair was labeled by a human expert, indicating the relevance of the result for the query. We also conduct experiments on 99 queries provided by the CodeSearchNet corpus for the Java programming language. We use the same metric, normalized discounted cumulative gain (NDCG~\cite{2008-Introduction-IR}), to evaluate baselines and {\toolname}. Our TranCS achieves NDCG of 0.223, outperforming DeepCS (\textbf{0.138}) and MMAN (0.173) by 62\% and 30\%, respectively.

\begin{table}[htbp]
  \renewcommand{\arraystretch}{1.0}
  \caption{Contribution of Key Components in {\toolname}}
  \label{tab:contribution_of_component}
  \centering
  \begin{tabular}{l|ccc|c}
    \toprule
     Tech & $\mathtt{SR@1}$ & $\mathtt{SR@5}$ & $\mathtt{SR@10}$ & $\mathtt{MRR}$\\
    \midrule
    TokeCS & 0.247 & 0.477 & 0.586 & 0.359 \\
    {\toolname} (CCT) & \textbf{0.352} & \textbf{0.569} & \textbf{0.664}  & \textbf{0.455} \\
    \midrule
    TokeCS (SWM) & 0.264 & 0.483 & 0.592 & 0.370 \\
    DeepCS (SWM) & 0.295 & 0.511 & 0.615 & 0.399 \\
    {\toolname} (CCT+SWM) & \textbf{0.560} & \textbf{0.764} & \textbf{0.824}  & \textbf{0.651} \\
    \bottomrule
 \end{tabular}
\end{table}

\subsubsection{\textbf{RQ2:} Contribution of Key Components}
\label{subsubsec:contribution_of_each_component}
We experimentally verified the effectiveness of two key components of {\toolname} i.e., context-aware code translation (CCT) and shared word mapping (SWM). 
In Table \ref{tab:contribution_of_component}, TranCS(CCT) and TranCS(CCT+SWM) are two special versions of TranCS, among which the former uses two different word mapping functions to transform instruction translations and comments to vector representations, while the latter uses SWM. 
In other words, if it is only CCT, {\toolname} uses two vocabularies. In the case of CCT+SWM, {\toolname} uses a shared vocabulary.  
Moreover, numerous existing studies \cite{2018-Neural-Code-Search, 2020-Code-Search-with-Co-Attentive-Representation, 2020-OCoR} including DeepCS \cite{2018-DeepCodeSearch} and MMAN \cite{2019-Multi-modal-Attention-for-Code-Rerieval} have shown that tokens of code snippets play a key role in code search tasks. 
Therefore, we assume that this is a scenario where the code snippet is not translated, and we directly pass the tokens of the code snippet into the model to train the code encoder. 
The effectiveness of the token-based CS technique (TokeCS) is shown in the second line of Table \ref{tab:contribution_of_component}. 
To demonstrate the effectiveness of SWM, we also tried to apply SWM to TokeCS, DeepCS and MMAN. 
To apply SWM to TokeCS, we use a unified word mapping function to transform tokens and comments. 
In DeepCS, the author uses four word mapping functions to transform the MN, APIS, Token and comments into vector representations. 
To apply SWM to DeepCS, we first merge the four vocabularies into a shared vocabulary by extracting the union of them. 
Then, we use a unified word mapping function to transform MN, APIS, Token and comments. 
In MMAN, the author not only uses LSTM architecture to embed tokens, but also uses Tree-LSTM and GGNN to embed AST and CFG, while the three architectures cannot share a word mapping function.
Therefore, SWM can not be applied to MMAN. 
The effectiveness of Toke(SWM), DeepCS(SWM) are shown in lines 4--5 of Table \ref{tab:contribution_of_component}. 
From the lines 2--3 of Table \ref{tab:contribution_of_component}, we can observe that for $\mathtt{SR@k}$, the improvements of TranCS(CCT) to TokeCS are 42.51\%, 19.29\% and 13.31\% when $k$ is 1, 5, and 10, respectively. 
For $\mathtt{MRR}$, the improvement to TokeCS is 26.74\%. 
Therefore, we can conclude that CCT contributes to {\toolname}. 
For $\mathtt{SR@k}$, the improvements of TranCS(CCT+SWM) to TranCS(CCT) are 59.09\%, 34.27\% and 24.10\%.
For $\mathtt{MRR}$, the improvement of TranCS(CCT+SWM) to TranCS(CCT) is 43.08\%. 
Therefore, we can conclude that SWM contributes to {\toolname}. 
Besides, we can also observe that SWM also has slight improvements to TokeCS and DeepCS. 
Therefore, we can draw the conclusion that SWM and CCT, which promote each other, improve the performance of {\toolname} jointly.

\subsubsection{\textbf{RQ3:} Robustness of {\toolname}}
\begin{figure}[htbp]
\graphicspath{{pictures/}}
\centering
    \subfigure[Varying Code Snippet Lengths]
    {
        \includegraphics[width=0.476\linewidth]{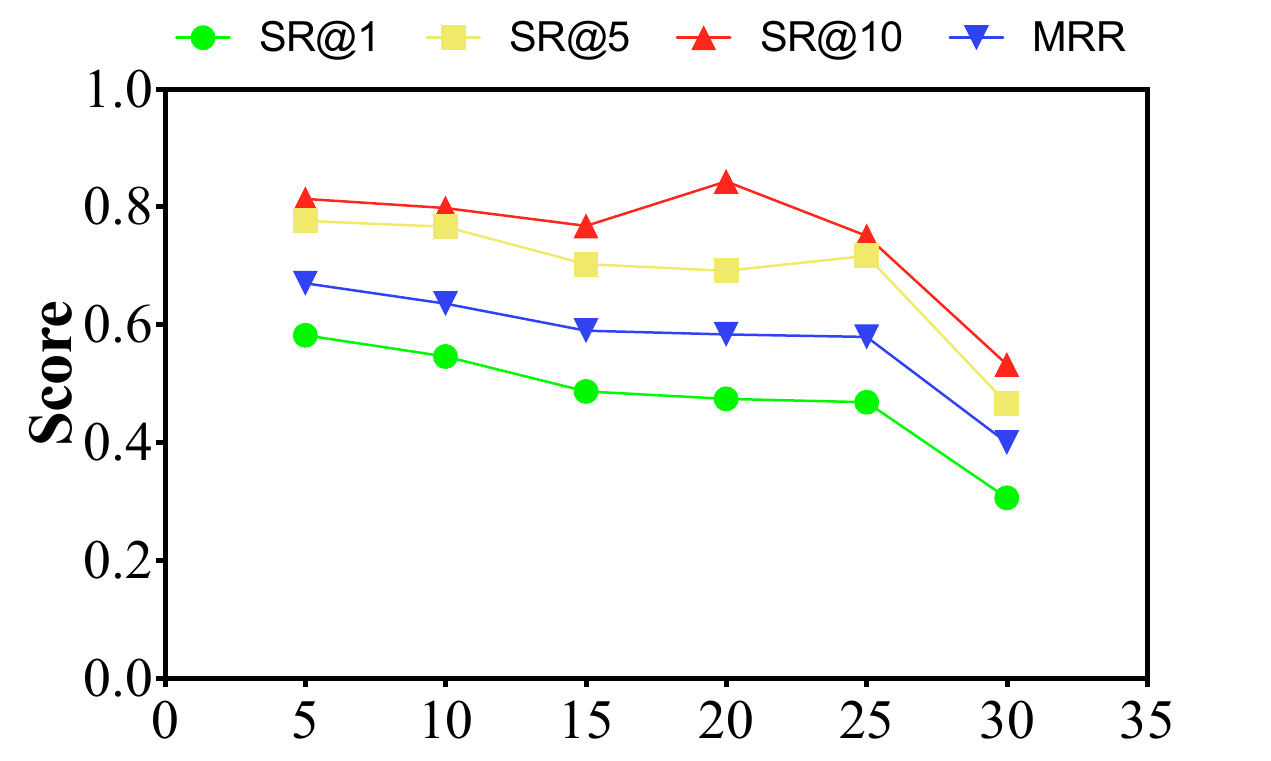}
        \label{fig:robustness_on_vary_code_length}
    }
    \subfigure[Varying Comment Lengths]
    {
        \includegraphics[width=0.476\linewidth]{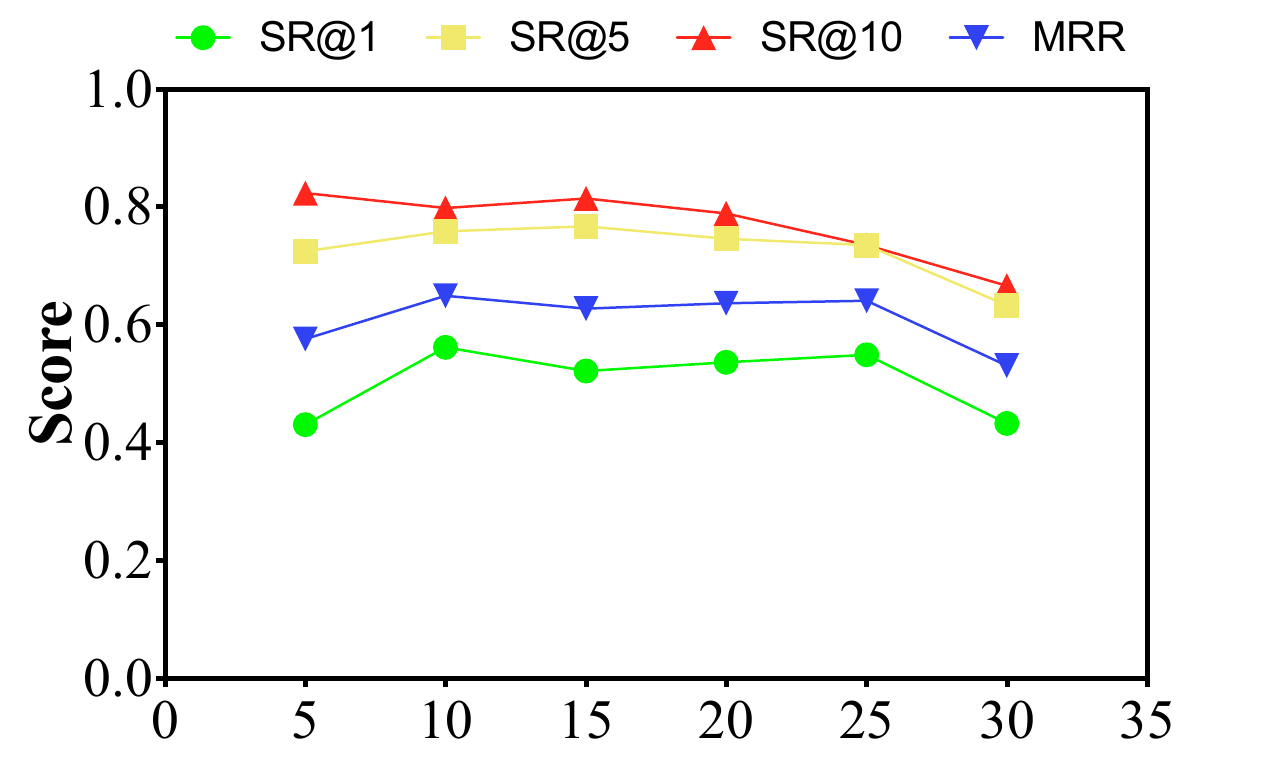}
        \label{fig:robustness_on_comment_length}
    }
    \caption{Robustness of {\toolname}}
    \label{fig:robustness_of_TranCS}
\end{figure}

\label{subsubsec:robustness_of_{\toolname}}
To analyze the robustness of {\toolname}, we studied two parameters (i.e., code length and comment length) that may have an impact on the embedding representations of translations and comments. 
Figure \ref{fig:robustness_of_TranCS} shows the performance of {\toolname} based on different evaluation metrics with varying parameters. 
From Figure \ref{fig:robustness_of_TranCS}, we can observe that in most cases, {\toolname} maintains a stable performance even though the code snippet length or comment length increases, which can be attributed to context-aware code translation and shared word mapping we proposed. 
When the length of the code snippet exceeds 20 (a common range described in Section \ref{subsubsec:dataset}), the performance of {\toolname} decreases as the length increases. 
It means that when the length of the code snippets or comments exceeds the common range, as the length continues to increase, it will be more difficult to capture their semantics. 
Overall, the results verify the robustness of our {\toolname}.

\section{Case Study}
\label{sec:qualitative_analysis}


\begin{figure}[htbp]
  \centering
  \includegraphics[width=\linewidth]{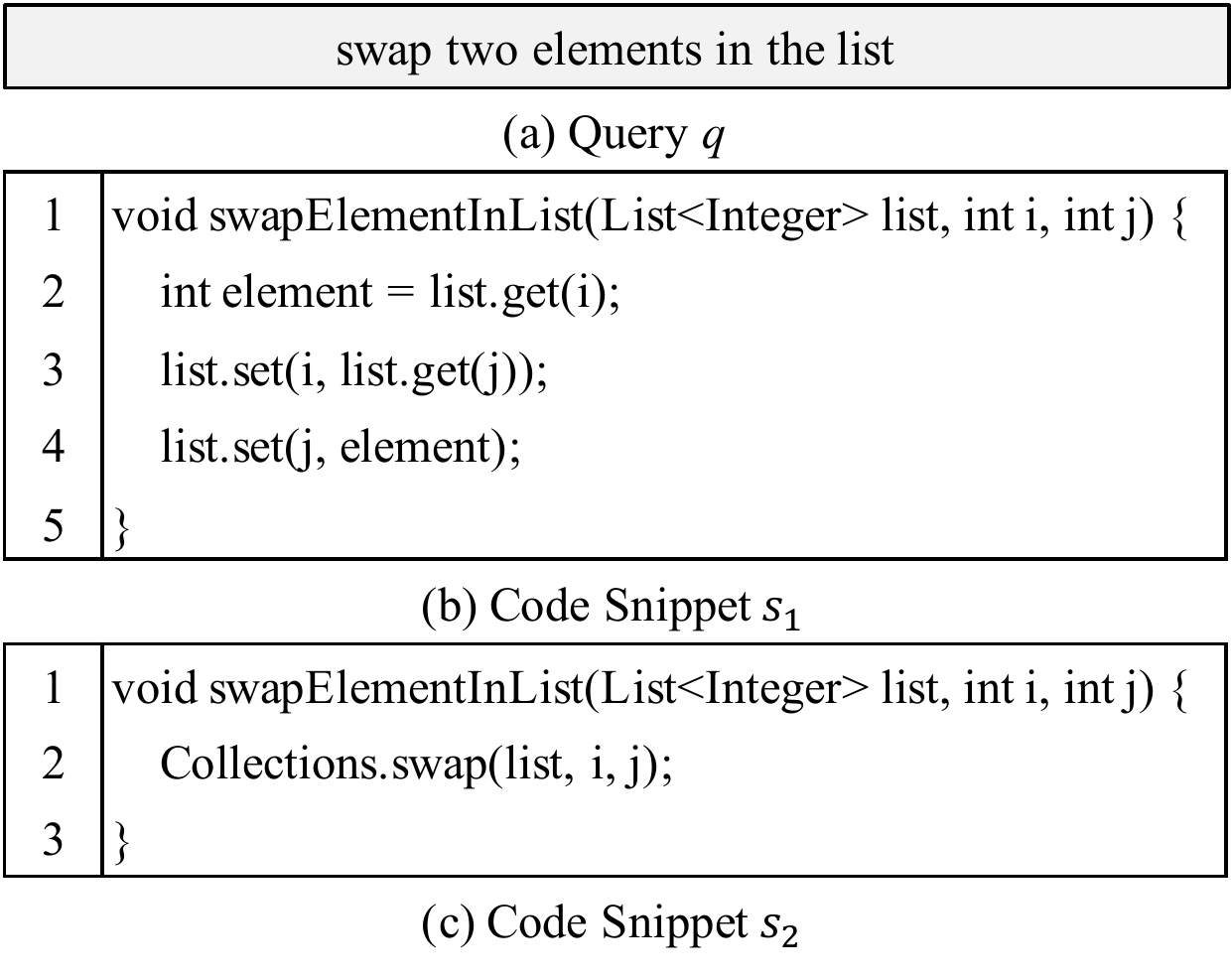}
  \caption{Example of Two Code Snippets Implementing the Same Functionality}
  \label{fig:example_of_case_I}
\end{figure}

\begin{figure}[htbp]
  \centering
  \includegraphics[width=\linewidth]{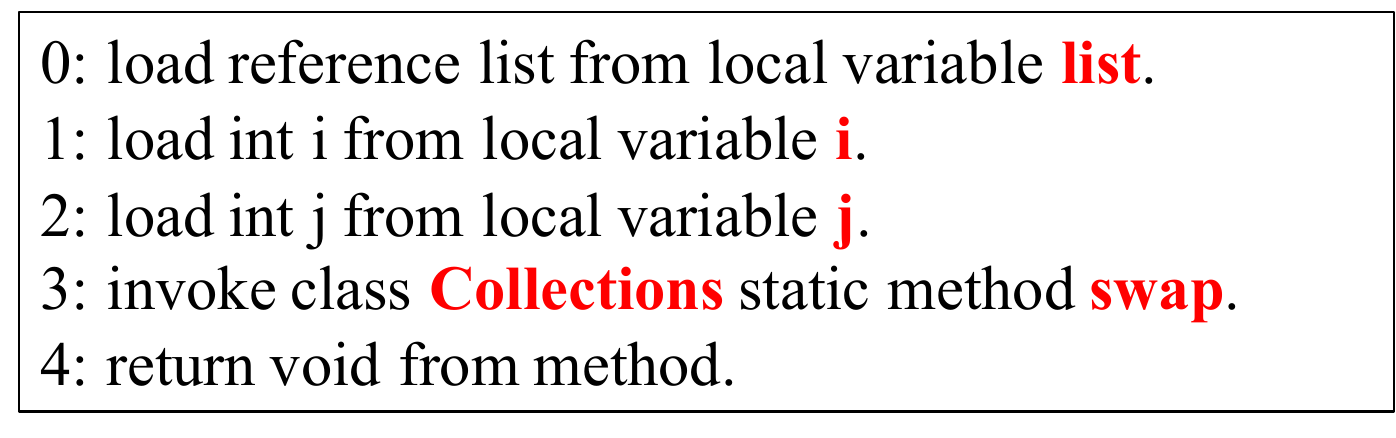}
  \caption{Translations of the Code Snippet $s_2$}
  \label{fig:case_of_implantation_difference}
\end{figure}

This is a case to study the performance of {\toolname} in retrieving code with implantation difference. Figure~\ref{fig:example_of_case_I}(b) and (c) show two code snippets that implement the same functionality, i.e., swapping two elements in the list.
The first one ($s_1$) implements the functionality from scratch, and the second one ($s_2$) directly calls the external API $Collection.swap()$. 
We use {\toolname} to convert the two code snippets into corresponding translations, which are very different, meaning {\toolname} can effectively differentiate semantically similar code but differs in APIs used. This is because {\toolname} reserves API information (e.g., name, parameter) when generating code translation. For example, as shown in Figure~\ref{fig:case_of_implantation_difference}, the translations produced by {\toolname} reserve the information of the API $Collection.swap()$ invoked by $s_2$, including the parameters (e.g., \textbf{list}) and the method name \textbf{swap}.  

\section{Threats to Validity}
\label{sec:threats_to_validity}
The metrics used in this paper are $\mathtt{SuccessRate@k}$ and $\mathtt{MRR}$ for evaluating the effectiveness of {\toolname} and existing techniques. These are the same metrics adopted in MMAN~\cite{2019-Multi-modal-Attention-for-Code-Rerieval}. We do not use another metric $\mathtt{Precision@k}$ that measures the percentage of relevant results in the top $k$ returned results for each query~\cite{2018-DeepCodeSearch}. This is due to the constraint that the relevant results need to be labelled manually, which is empirically less feasible and can introduce human biases. We hence focus on the two metrics $\mathtt{SuccessRate@k}$ and $\mathtt{MRR}$ in the paper.

{\toolname} is currently only evaluated on Java programs and may require modifications for extending to other programming languages. The core contribution of {\toolname} is the context-aware code translation technique. To realize the context-aware code translation, {\toolname} requires a set of translation rules, such as the operations and descriptions of instructions. In order to extend {\toolname} to other programming languages, corresponding translation rules need to be designed and provided. We plan to evaluate the performance of {\toolname} on these programming languages in future work. 

\section{Related Work}
\label{sec:related_work}
Early CS techniques were based on IR technology, such as~\cite{2006-Source-Code-Exploration, 2010-Example-centric-programming, 2011-Portfolio, 2014-Spotting-Working-Code, 2015-How-Developers-Search-for-Code}. 
These techniques simply consider queries and code snippets as plain text and then use keyword matching. 
To alleviate the problem of keyword mismatch~\cite{2009-Automatically-Capture-Code-Context, 2012-Survey-of-Automatic-Query-Expansion} and noisy keywords~\cite{2016-Deep-API-Learning}, many query reformulation(QR)-based CS techniques~\cite{2009-Automatically-Capture-Code-Context, 2014-Thesaurus-based-Query-Expansion, 2015-CodeHow, 2015-Query-Expansion-via-WordNet, 2016-Query-Expansion-based-Crowd-Knowledge, 2018-Effective-Reformulation-of-Query} have been proposed one after another. 
For example, the words from WordNet~\cite{1995-WordNet}, or Stack Overflow~\cite{2016-Query-Expansion-based-Crowd-Knowledge} are used to expand user queries.
However, QR-based CS techniques consider each word independently, while ignoring the context of the word. 
In addition, both IR-based and QR-based CS techniques only treat the code snippet as plain text, and cannot capture the deep semantics of the code snippet. 
To better capture the semantics of queries and code snippets, deep learning (DL)-based CS techniques~\cite{2018-DeepCodeSearch, 2018-Neural-Code-Search, 2019-Multi-modal-Attention-for-Code-Rerieval, 2019-DL-Met-CodeSearch, 2019-Coacor, 2020-Code-Search-with-Co-Attentive-Representation, 2020-CodeBERT, 2019-CodeSearchNet-Challenge} have been proposed one after another. 
Gu et al.~\cite{2018-DeepCodeSearch} first apply DL to the code search task. 
They first encode both the query and a set of code snippets into corresponding embeddings using MLP or RNN, and then rank the code snippets according to the cosine similarity of embeddings. 
Other DL-based CS techniques are similar to DeepCS~\cite{2018-DeepCodeSearch} with only a difference in choosing the embedding architecture. 
For example, to capture the semantics of other aspects of the code snippet, MMAN~\cite{2019-Multi-modal-Attention-for-Code-Rerieval} integrates multiple embedding networks (i.e., LSTM, Tree-LSTM and GGNN) to capture semantics of multiple aspects, such as Token, AST, and CFG. 
CodeBERT~\cite{2020-CodeBERT}, CoaCor~\cite{2019-Coacor}, and baselines in CodeSearchNet Challenge~\cite{2019-CodeSearchNet-Challenge} only treat the code snippet as plain text (token sequence), which miss richer information such as APIs, AST, and CFG, etc. 
TBCNN~\cite{2016-TBCNN} is a tree-based convolutional neural network that encodes the AST of the code snippet. Our baseline MMAN has encoded AST using tree-based neural networks and is inferior to our {\toolname}. 
All these works have a similar idea that first transforms both code snippets and queries into embedding representations into a unified embedding space with two encoders, and then measures the cosine similarity of these embedding representations. 
However, {\toolname} differs from previous work in two major dimensions: 
1) {\toolname} first translates the code snippet into semantic-preserving natural language descriptions. 
In this case, the generated translations and comments are homogeneous. 
2) Based on code translation, {\toolname} naturally uses a shared word mapping mechanism, which can produce consistent embeddings for the same words, thereby better capturing the shared semantic information of translations and comments.

\section{Conclusion}
\label{sec:conclusion}
In this paper, we propose a context-aware code translation technique, which can translate code snippets into natural language descriptions with preserved semantics. 
In addition, we propose a shared word mapping mechanism to produce consistent embeddings for the same words/tokens in comments and code snippets, so as to capture the shared semantic information. 
On the basis of context-aware code translation and shared word mapping, we implement a novel code search technique {\toolname}. 
We conduct comprehensive experiments to evaluate the effectiveness of {\toolname}, and experimental results show that {\toolname} is an effective CS technique and substantially outperforms the state-of-the-art techniques. 

In future work, we will further explore the following two dimensions: (1) as shown in Figure\ref{fig:distribution_of_length}, statistical results on large-scale data sets show that most code snippets have no more than 20 lines. Within this range, {\toolname} is robust and stable. Constructing representations of long code snippets is still an open problem, and we leave it to future work.
(2) LSTM encoder is just a component of {\toolname}, which can be easily replaced with more advanced (including pre-trained) models in~\cite{2021-Corder, 2021-Studying-T5-Support-Code-Related-Tasks}. We will explore more advanced models in future work.

\section*{Acknowledgement}
The authors would like to thank the anonymous reviewers for insightful comments. This work is supported partially by National Natural Science Foundation of China(61690201, 62141215).

\bibliographystyle{ACM-Reference-Format}
\bibliography{reference}

\end{document}